\documentclass[fleqn,usenatbib]{mnras}

\usepackage{float}
\usepackage{longtable}
\usepackage{subcaption}
\usepackage{mathptmx}
\usepackage{siunitx}
\usepackage{arydshln}

\usepackage[T1]{fontenc}
\usepackage{ae,aecompl}
\usepackage{color}
\usepackage{enumerate}

\usepackage{graphicx}	
\usepackage{amsmath}	
\usepackage{amssymb}	
\usepackage{hhline}
\usepackage{hyperref}

\definecolor{darkred}{rgb}{0.8, 0.0, 0.0}
\definecolor{navy}{rgb}{0.2, 0.2, 0.7}
\definecolor{fred}{rgb}{0.5, 0.3, 0.7}

\title[Improved selection of ERQs]{Improved selection of extremely red quasars with boxy CIV lines in BOSS}

\author[R. Monadi \& S. Bird]{
Reza Monadi\thanks{E-mail: reza.monadi@email.ucr.edu } and
Simeon Bird \thanks{E-mail: sbird@ucr.edu }
\\
University of California, Riverside, 92507 CA, U.S.A.}

\date{Accepted 2022 January 27. Received 2021 November 22; in original form 2021 August 2
}

\pubyear{2020}
\newcommand{\civ}{\ion{C}{IV}}
\newcommand{\nv}{\ion{N}{V}}
\newcommand{\fnc}{\nv/\civ}
\begin{document}
\label{firstpage}
\pagerange{\pageref{firstpage}--\pageref{lastpage}}
\maketitle

 \begin{abstract}
Extremely red quasars (ERQs) are an interesting sample of quasars in the Baryon Oscillation Spectroscopic Sample (BOSS)
in the redshift range of $2.0 - 3.4$ and have extreme red colours of $i-W3\ge4.6$.  
Core ERQs have strong \civ~emission lines with
rest equivalent width of $\ge100$\AA. Many core ERQs also have \civ~line profiles
with peculiar boxy shapes which distinguish them from normal blue quasars.
We show, using a combination of kernel density estimation and local 
outlier factor analyses on a space of the $i-W3$ colour, \civ~rest 
equivalent width and line kurtosis, that core ERQs likely represent a
 separate population rather than a smooth transition between normal 
 blue quasars and the quasars in the tail of the colour-REW distribution.
We apply our analyses 
to find new criteria for selecting 
ERQs in this 3D parameter space. Our final selection 
produces $133$ quasars, which are \emph{three} times more 
likely to have a visually verified \civ~broad absorption 
line feature than the previous core ERQ sample. We further show that
our newly selected sample are extreme objects in the
intersection of the WISE AGN catalogue with the
MILLIQUAS quasar catalogue in the colour-colour space of ($W1-W2$, $W2-W3$).
This paper validates an improved selection method for red quasars which
can be applied to future datasets such as the quasar catalogue from the
Dark Energy Spectroscopic Instrument (DESI).

\end{abstract}




\section{Introduction}

Quasars are high luminosity active galactic nuclei (AGN), fuelled by
gas and dust accreting onto a supermassive black hole (SMBH). Observations show that
the growth of a SMBH is correlated with the physical properties of the host galaxy, such as velocity dispersion, stellar mass, and star formation rate, although the mechanisms which correlate these properties are not completely clear
  \citep{gebhardt-2000, tremaine-2002,
m-sigma,gultekin-2009, shankar-2009, kormendy-2013, azadi-2015, graham-2016}.
The co-evolution of a SMBH and its host galaxy mostly occurs during dusty starbursts
resulting in the observation of sub-mm or ultra-luminous infrared
 galaxies \citep{sanders-1988, veilux-2009, simpson-2014}.
 \cite{Pavel2020} proposed a model whereby SMBHs form when the dynamical collapse of 
 star clusters is accelerated by the accretion of gas from a host galaxy, which can naturally 
 explain the correlation between a host galaxy and SMBH mass, as well as explain quasars found at high redshift.

Unobscured  quasars which exhibit blue thermal continuum are the
majority of optically selected quasars. \emph{Red quasars}, on the other hand,
are a small population of quasars that
show a variety  of redder near infra-red and optical colours.
Several studies have investigated the origin of the red colour in red quasars
(eg. \cite{kim-2018} or \cite{radio-red-diff-1}), however
the question still remains unsettled \citep{rivera-2021}.

One possibility is that red quasars have been obscured and reddened by dust during
a brief transition phase between dusty starburst galaxies and blue quasars
\citep[e.g.~][]{richards-2003, hopkins-2005, urrutia-2008, hopkins-2008, glikman-2012,
glikman-2015, banerji-2015, assef-2015, ishibashi-2016, hickox-2018}.
In this model, a quasar is buried in the
starburst dust when the host galaxy is young, making the colour of that
quasar red.
Many ERQs also exhibit extreme line properties
which may indicate unusually powerful outflows occurring in a young evolution phase.
Quasar-driven outflows may clear out the observer's line of sight, and so at the
end of this evolutionary phase, we observe an optical and/or UV luminous
quasar.

There are other models; for example, the unified AGN model \citep{unified-agn}
suggests that red quasars are viewed with intermediate orientations
between Type 1 and Type 2 quasars (for a recent review see \cite{hickox-2018}).
According to this model, unobscured (type 1) AGN are viewed face-on, while
obscured (type 2) AGN are observed edge-on.
A red colour is produced by a dusty torus blocking part of the
nuclear emission. However, this model has difficulty explaining the
extreme line properties seen in some red quasars \citep{Urrutia-2009, radio-red-diff-1}.


\cite{ross-15}  studied a population of red quasars at $0.28\le z \le 4.36$
in the
Baryon Oscillation Spectroscopic Survey \citep[BOSS][]{boss} of the Sloan Digital Sky
Survey-III \citep[SDSS-III][]{sdss3}. These red quasars were identified using
a simple colour selection originally intended for red galaxies: a magnitude difference
of $r_{AB} - W4_{Vega} \ge 14$ between the infra-red
band ($W4$ in WISE with effective wavelength of $12\mu$m)
and the optical band ($r$ in SDSS with effective wavelength of $6231$\AA).


\cite{hamann17} (hereafter H17) used the sample of \cite{ross-15} but narrowed down
the redshift range to $2.0 \le z \le 3.4$ and
changed the colour selection to $i-W3 \ge  4.6$ ($\sim 3$ magnitudes redder than
 the typical colour of BOSS quasars), calling the sample thus identified 
 Extremely Red Quasars, or ERQs\footnote{ERQs are not the reddest $i-W3$
 quasars overall, but the reddest ones in BOSS.}. Interestingly, ERQs showed exotic spectral
  properties, which motivated H17 to define a smaller core ERQ (CERQ)
   subsample defined by REW(\civ)$\ge100$\AA.
This criterion was chosen to better correlate red quasars with other extreme
line properties: peculiar \emph{boxy} profiles, N{\tiny{V}}$>$Ly$\alpha$, a high
incidence of blue-shifted broad absorption lines (BALs), and
[OIII] 5007\AA \ outflow speeds reaching $>6000$ km/s.
ERQs also have an unusually flat UV SED
considering their extreme red colour (steep Mid-IR to UV SED), although this may be an artifact of the
BOSS selection algorithm, which would not target quasars which are red in
all SDSS bands for spectroscopic follow-up.

\cite{serena19} (hereafter P19) studied a sample of 28 ERQs and found
an outflow speed for the [OIII] line of $1992$ - $6702$ km/s.
This is on average three times faster than those of luminosity
matched blue quasars. This outflow speed is highly
correlated with $i-W3$ colour but not with radio loudness
nor Eddington ratios. P19 suggests that this correlation may indicate
a connection between reddening and the efficiency of energy and momentum
injection from ERQs to the interstellar medium.
ERQs may produce more effective feedback in
their host galaxies, regulating the star formation
rate and SMBH growth more effectively. This is again
indicative that some ERQs with extreme line values are
connected with an early dusty stage of quasar-galaxy
evolution where strong quasar-driven outflows provide
important feedback to the host galaxies (P19).

We therefore have a working hypothesis identifying ERQs with an intermediate
stage of quasar evolution between dusty galaxies and red quasars.
This study is an effort to produce quantitative evidence
for or against this hypothesis, which is based on the unusual line properties
exhibited by the spectra. If such quantitative evidence is forthcoming, we also
desire to refine the selection criteria for ERQs in order to better study the
outflows connected with this stage of quasar evolution. 
We will provide selection criteria for objects that exhibit the
extreme properties of ERQs, among a sample of quasars with spectroscopic data.
We use the existing manually selected sample of ERQs
to define a training set, and then provide a modified sample
of extremely red quasars in BOSS with more uniform (and more
uniformly exotic) properties.
In summary, we address the following questions:
\begin{enumerate}[1)]
\item To what extent are ERQs separated
from the main locus of BOSS quasars (\S\ref{sec:density}, \S\ref{sec:med-spec} and \S\ref{sec:lof})?
\item If they are, which selection criteria best produce quasars
 connected with this intermediate stage of quasar 
evolution \S\ref{sec:3d-boundary})?
\end{enumerate}
We acknowledge the possibility that our sample may be affected by the selection criteria of BOSS, which uses a colour selection to find quasar candidates and thus may discard some red quasars. However, in the absence of another equally large spectroscopic quasar survey this is unavoidable. We will thus analyse BOSS quasars and check for evidence that we are affected by colour selection in Section \ref{sec:milliquas}.

We use a standard cosmology throughout
(H$_{0} = 67.3$ km s$^{-1}$ Mpc $^{-1}$, $\Omega_{m}^0 = 0.315$,  $\Omega_{\Lambda}=0.685$) \citep{planck}.

\section{Quasar samples }\label{sec:samples}

\begin{table}
	\centering
   \caption{The parent samples, sizes and sample selection criteria for the various quasar samples we use. All subsets are taken from a parent sample made in H17 by custom emission line fits. The first sample, T1, is a superset of all the others. These are: type 1 luminosity matched quasars (T1LM),
          type 1 extremely red quasars (T1ERQ) and type 1 core extremely
          red quasars (T1CERQ).}
  \label{tab:selection}
	\begin{center}
	\begin{tabular}{|l|c|l|}
		\hline
		Sample  & Selection criteria & Size \\  \hline
		T1 &  \shortstack{FWHM(\civ) $\ge$ 2000 km.s$^{-1}$ \\
		    2 $\le$ z\_dr12 $\le$ 3.4 \\
			$i - W3 \ge0.8$ \\ SNR(REW(\civ)$\ge$3 \\
			SNR(FWHM(\civ)$\ge$4 \\
			SNR(AB$_{W3})\ge$ 3	\\\texttt{q\_flag =0} \\\texttt{cc\_flag='0000'} \\\texttt{nv\_flag=0}} & 35,976 \\ \hline
        T1LM  & $10^{46.54}$ erg.s$^{-1}$  $\le$ L$_{bol}\le 10^{48.00}$ erg.s$^{-1}$  &  29,072\\ \hline
		T1ERQ &  i-W3 $\ge 4.6$ & 154 \\ \hline
      T1CERQ & \shortstack{i-W3 $\ge 4.6$ \\ REW(C{\tiny{IV}})$\ge$ 100\AA} & 72 \\	\hline
	\end{tabular}
	\end{center}
\end{table}
In this section, we introduce our quasar samples their selection
criteria, summarised in Table \ref{tab:selection}.
The primary parent sample is similar to the emission-line catalogue
of H17 which results from custom fits of \civ \ and \nv \ emission
lines performed on spectra in the SDSS-III BOSS quasar catalogue,
The subsamples follow the selections in H17, to which we refer
the reader for a detailed explanation of the criteria adopted.

\subsubsection{Type 1 sample}

Following the H17 sample selection procedure, we first limit the
quasar redshift to $2 \leq z \leq 3.4$. This redshift range encompasses most of the
BOSS survey, while ensuring that Ly$\alpha$ and \ion{N}{V} $\lambda$1240
are within the BOSS spectral range.
We require that successful fits to the \ion{N}{V} (\texttt{nv\_flag=0}) 
and \civ~(\texttt{q\_flag =0}) emission lines are made at reasonable signal
 to noise (SNR(REW(\civ)) $\ge 3$ and SNR(FWHM(\civ)$\ge 4$)). We limit ourselves
  to Type 1 quasars, defined as FWHM(\civ) $\ge 2000$ km s$^{-1}$ \citep{alex-2013,ross-15}.
We also require that the quasars have a good detection in the W3 
band (SNR(AB$_{W3})\ge 3$), do not exhibit artifacts in the WISE data (\texttt{cc\_flag='0000'}),
and are not excessively blue ($i - W3 > 0.8$).

%

\subsubsection{T1ERQ and T1CERQ samples}

Following H17 we use a colour cut of $i - W3 \ge 4.6$ to extract ERQs from the
full sample of type 1 quasars. H17 considered several colour cuts, choosing
their boundary to produce the most dramatic differences in the median spectral
properties of ERQs as compared to blue quasars. H17 also
 defined a subsample, core type 1 ERQs (T1CERQs), with the
  additional condition of REW(\civ)$\ge 100$\AA, chosen to be more correlated
   with the unusual line properties found in some ERQs.
    These conditions define a natural two dimensional
     parameter space in $i-W3$ and REW(\civ), which we will use extensively in what follows.





\subsubsection{T1LM sample}
ERQs are very luminous, with an average bolometric luminosity for
 T1CERQs of $10^{47.21\pm0.31}$ erg.s$^{-1}$. For comparison, the
full quasar sample has an average luminosity of $10^{46.82\pm0.21}$ erg.s$^{-1}$.
This high luminosity is a selection effect. SDSS cannot detect faint ERQs, because
it does not detect objects with an $i$ band magnitude $\lesssim 22$.


This large luminosity has implications for our later analysis, as there is an
anti-correlation between the REW of the \civ~emission line and the continuum luminosity
of Type 1 quasars \citep{Baldwin}. We made a sample of Type 1 luminosity 
matched quasars (T1LM) drawn from the \emph{T1 sample}, but with luminosities
 between the minimum and maximum luminosities of T1CERQs (ie. $10^{46.54}\le L_{bol}(T1CERQ)\le 10^{48.00}$).

We derive mean luminosities using the procedure described in P19, using the AB
 magnitude in the W3 band and the luminosity distance of our standard cosmology.

Bolometric luminosities are difficult to determine for
 ERQs due to the large and uncertain extinction corrections which must be
applied in the rest-frame UV and optical. However, the W3 band is less susceptible
to obscuration than the UV and optical bands. We therefore use it as a surrogate for estimating
the bolometric luminosity. We add a correction for flux suppression due to obscuration of a factor of $8$, following H17. We thus set $L_\mathrm{bol}=8\lambda L_\lambda$ at $\lambda = 3.45 \mu$m in the rest frame.
  WISE W3 photometry measures $\sim 3.45\mu$ m in the rest-frame at the typical
   redshift of ERQs in our study.
   We first converted the AB magnitude of W3 to units of flux per frequency
($F_{\nu} = 10^{-0.4(48.6+m_{AB})}$) and then
to observed flux per wavelength ($F_{\lambda} = \frac{c}{\lambda^2}F_{\nu}$).
Intrinsic luminosity is computed using the luminosity distance, $D_L$, of our
 standard cosmology by:
\begin{equation}
 L_{\lambda} \lambda_{rest} = 4\pi D_L^2 F_{\lambda} \lambda_{obs}
\end{equation} 

Even after luminosity matching, the sample may contain quasars with a range of different
black hole masses or Eddington ratios. However, reliable measures of these quantities
are not generally available in our dataset, and so to avoid the risk of over-fitting
we rely on simple luminosity matching to homogenise our sample.

\section{Analysis Methods}
\label{sec:analaysis}

\subsection{Kurtosis of the CIV line: a third parameter}
\label{sec:distk80}

H17 investigated a large number of unusual emission line properties in the T1ERQ and T1CERQ samples.
In particular, they found boxy \civ~line shapes, a large \nv \ to \civ \ line ratio (\fnc) and
moderately reduced FWHM(\civ) compared to a population of normal
blue quasars with similar W3 magnitudes to T1CERQs. However, the \ion{N}{v} fits done in H17
did not attempt to deblend the nearby Lyman-$\alpha$ line and so the \ion{N}{v} strength may be overestimated. The \ion{C}{IV} line is uniquely powerful for our analysis as it is the strongest metal line in the quasar spectrum. Other promising lines (e.g.~\ion{Si}{IV} or \ion{He}{II}) lines are much weakeror blended. For example, \ion{C}{III} is blended with \ion{S}{III} and \ion{Al}{III}.

We focus here on the boxy shape of the \civ~line, quantified by the kurtosis.
 Kurtosis (kt$_{80}$) is defined in H17 as the ratio of the velocity width of
  the \civ \ line at 80\% of the peak height to the velocity width at 20\% of the peak height.
A high kurtosis \civ \ line profile occurs in most ERQs and indicates a boxy line.
The median kt$_{80}$ for T1ERQs is $0.35$ and for T1CERQs $0.36$, while the
larger T1LM sample has a median of $0.25$.





\begin{figure*}
    \centering
\includegraphics[width=\linewidth]{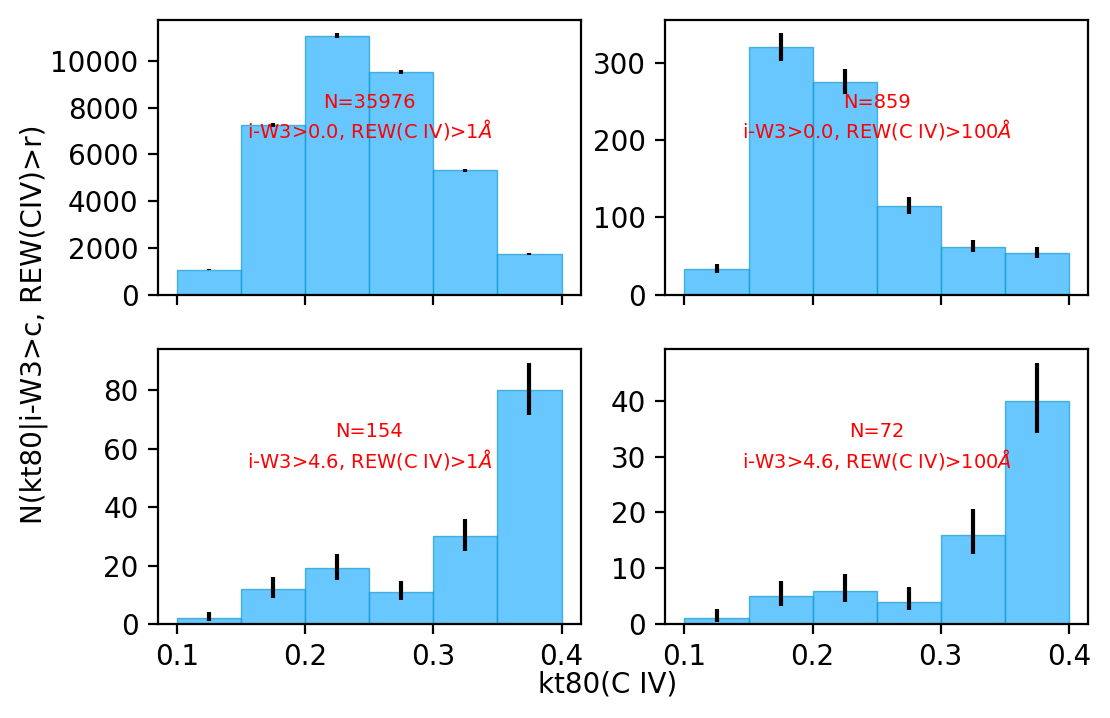}
\caption{Histograms for distribution of kt$_{80}$(\civ) given
 conditions on $i-W3$ and REW(\civ). Top left panel shows the unconditioned distribution.
Bottom left panel is conditioned on $i-W3 > 4.6$, and thus shows T1ERQs. 
Top right panel is conditioned on REW(\civ)$>100$\AA~and bottom right panel 
is conditioned on both, thus showing T1CERQs.
Each panel is labelled by the conditions and the number of quasars
 satisfying them. $c$ and $r$ in $N(kt_{80}(\civ)|i-W3>c, \ REW(\civ)>r)$ are the colour and REW(\civ) thresholds shown in each panel.}
\label{matrix-hist}
\end{figure*}

Figure~\ref{matrix-hist} show histograms of kt$_{80}$(\civ). Each panel
is labelled by joint thresholds on REW(\civ) and $i-W3$ colour
and the number of quasars satisfying these conditions.
A redder colour skews the kt$_{80}$(\civ) distribution towards more
boxy \civ \ line quasars (higher kt$_{80}$(\civ)).
Increasing
the REW(\civ) threshold does not change the overall shape of
the kt$_{80}$(\civ) distribution dramatically, nor its most probable
value when compared to the unconditioned sample in the top left panel. However,
we see a slightly enhanced population of high kurtosis objects when
 conditioning on REW(\civ). High kt$_{80}$(\civ) is thus highly
  correlated with red colour, but not with
REW(\civ), suggesting that it is a good choice for a third parameter, along with
$i-W3$ and REW(\civ).

\begin{figure}
\includegraphics[width= \linewidth]{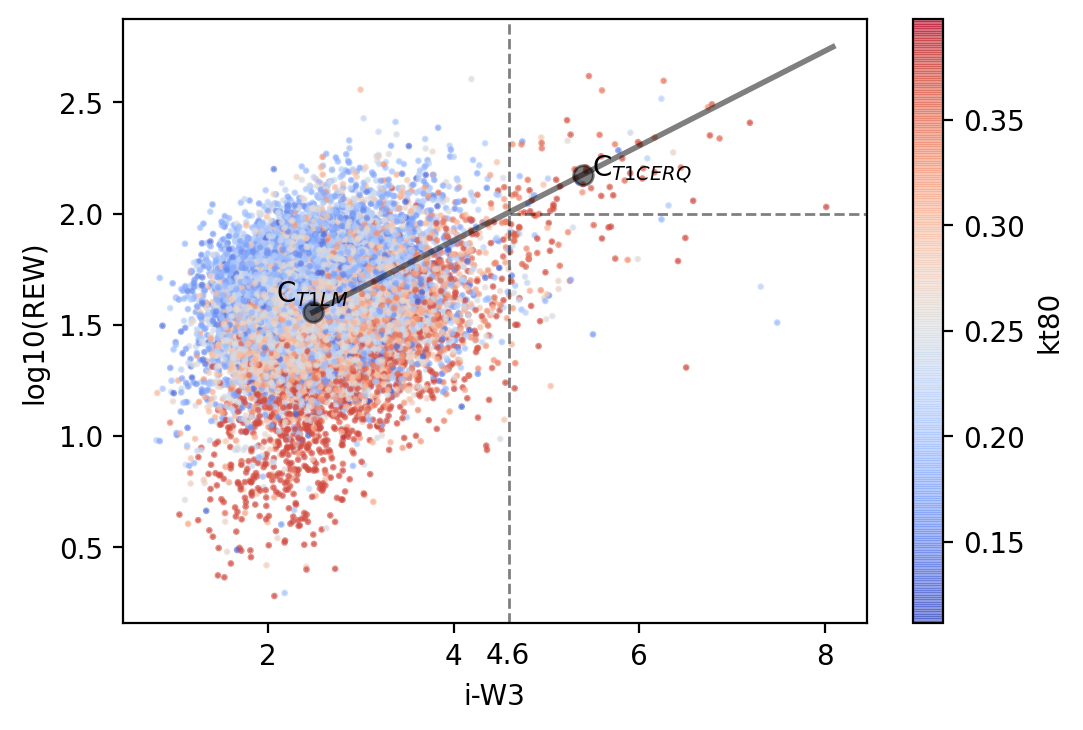}
\caption{Luminosity matched sample distribution in ($i-W3$, REW(\civ), kt$_{80}$(\civ))
space. Redder points show higher kt$_{80}$(\civ) and thus higher kurtosis.
The vertical line separates the T1ERQ sample
from the rest of T1LM sample and the horizontal line separates T1CERQs
from other T1ERQs. The black line is along $\vec{v}_{T1CERQ}$ (see \S \ref{sec:wedge-cone}), which connects  the median of the T1LM ($C_{T1LM}$) sample
to the median of the T1CERQ  ($C_{T1CERQ}$) sample.}
\label{kt80-cmap}
\end{figure}

Note that there is a possible confounder in the fitting procedure
of H17: weak lines will be fit with a single Gaussian rather than two
if the second Gaussian does not improve the fit. A single Gaussian
has $kt_{80}=0.37$. We have checked that this does not significantly
affect our results by making a version of Figure~\ref{matrix-hist}
where spectra with kt$_{80}$(\civ) $ > 0.37$ have been removed.
This reduces the total size of the sample by $34780$ spectra and
the number of core ERQs by $57$. In practice, since most of the
removed spectra are not ERQs this cut moderately strengthens
the trends we report. Fig. \ref{kt80-cmap} shows these high  kt$_{80}$(\civ) 
objects in the low REW(\civ) and blue part of the parameter space. 
The median REW(\civ) for kt$_{80}$(\civ) $ > 0.37$
and kt$_{80}$(\civ) $ > 0.36$ are 17\AA\ and 19\AA\ respectively. This indicates
that most of the quasars with high kt$_{80}$
are weak \civ~line objects, very far from the ERQs in colour space.



\subsection{Defining T1CERQs with a wedge or a cone}
\label{sec:wedge-cone}
One of our main goals in this study is to examine  variations in quasar spectral
properties as one moves in parameter space between the main quasar locus and T1CERQs.
We define $\vec{v}_{T1CERQ}$, the vector in the 2D parameter space of
colour-REW(\civ) between the median of the T1LM sample and the median of the T1CERQ sample, where the black line in Figure~\ref{kt80-cmap} shows its direction.
In order to average over quasar properties, we then define wedges (in 2D) and cones (in 3D), the simplest directional geometric shapes for calculating
the median spectra over a region in parameter space. We will use these shapes extensively in the following analysis\footnote{It is possible to consider more complex geometries. However
this makes the analysis over-complicated and is not necessarily better than a 
wedge (or cone) which covers an area (or volume), as long as a variety of directions are included.}.


A complexity to our definitions of wedges and cones is that we need a dimensionless parameter space within which to define opening angles.
We thus normalise all parameters to a dimensionless unit square (cube), based on the range of
each parameter.
The normalization procedure we choose is the min-max method,\footnote{We used the \texttt{MinMaxScaler} function from
\href{https://scikit-learn.org/stable/modules/generated/sklearn.preprocessing.MinMaxScaler.html}{\texttt{sklearn}}}
which performs a linear transformation to map each coordinate onto a unit square (cube).
The maximum value found in the dataset maps to $1$ and the minimum maps to $0$.
For our dataset, $i-W3$ ranges between $0.8$ and $8.0$, $\log10$(REW(\civ)) between
$0.2$ and $2.6$, and kt$_{80}$(\civ) between $0.11$ and $0.39$.

We define a wedge in a 2D space of $i-W3$ and REW(\civ) along the vector between
the median of the T1LM quasar sample and the T1CERQ sample. This vector is
$\vec{v}^{2D}_{T1CERQ}=(0.40, 0.26)$ in the normalised space, which corresponds to
$(2.90,136$\AA$)$ once the normalisation is removed. The opening angle for
this wedge is calculated by:
\begin{equation}\label{theta}
\theta = \max_i \{ \measuredangle \vec{P_i}, \vec{v}_{T1CERQ}\}.
\end{equation}
$\vec{P_i}$ is a vector
from the median of the T1LM sample and the $i$th quasar in the T1CERQ sample.
$\measuredangle$ means the angle between two vectors and is always less than
$180^{\degr}$. Eq. \ref{theta} implies that $\theta$ is the maximum angle
among all deviation angles of T1CERQs from $\vec{v}_{T1CERQ}$.
$\theta$ is thus the smallest angle for which the wedge
covers the entire T1CERQ sample,
quasars with $i-W3\ge4.6$ and REW(\civ)$\ge 100$\AA.
The opening angle in the \emph{normalised} 2D parameter space of $i-W3$ and
REW(\civ) for  $\vec{v}_{T1CERQ}^{2D}$ is
$\theta=18.5^{\circ}$.

Equivalently, in the 3D normalised space of 
($i-W3$, REW(\civ), kt$_{80}$(\civ))
we defined a cone towards the median of  T1CERQs.
But we observed that spectra with low kt80 did not generally
exhibit the exotic line properties of T1CERQs. 
Therefore, we impose a minimum kt$_{80}$(\civ) $\ge 0.33$.
This cut excludes  quasars with low kt$_{80}$(\civ) which have large 
$\theta$ angles from  $\vec{v}_{T1CERQ}^{3D}$ and  keeps the cone  
 focused on the  $\vec{v}_{T1CERQ}^{3D}$ direction.
This threshold is chosen because it corresponds to the dip in the
population distribution of kt$_{80}$(\civ) conditioned on red colour and
strong REW(\civ) (bottom right panel of Figure~\ref{matrix-hist}).
If we do not impose kt$_{80}$(\civ)$\ge 0.33$, the cone
opening angle will be large, $35.7^{\circ}$, and will
include many interloper quasars without the extreme line
properties of T1CERQs.
The corresponding vector between the median of the T1LM quasar sample and the
T1CERQ sample is $\vec{v}_{T1CERQ}^{3D}=(0.40, 0.26, 0.38)$, which is
$(2.90, 136$\AA$, 0.11)$ when the normalisation is removed.

The 3D cone with this definition thus includes all
quasars with $i-W3\ge4.6$, REW(\civ)$\ge100$\AA, \ and kt$_{80}$(\civ)$\ge 0.33$.
It has an opening angle of $\theta=19.6^{\circ}$ in the normalised 3D space.

Figure~\ref{kt80-cmap} visualises the boundaries between the T1LM, T1ERQ,
and T1CERQ quasar samples in the parameter space of $i-W3$ and REW(\civ).
Colours denote $kt_{80}$, showing again the large $kt_{80}$ associated with ERQs.
The line in Figure \ref{kt80-cmap} is along $\vec{v}_{T1CERQ}$, directed
from the median of the T1LM sample to the median of the T1CERQ sample.

\subsection{Local Outlier Factor Analysis}\label{sec:mocklof}

Wishing to investigate whether T1CERQs are a separate sub population of
 quasars, we applied several clustering methods on our dataset. We tried density-based clustering techniques
\cite[e.g.~DBSCAN][]{dbscan} and hierarchical clustering algorithms \cite[e.g.~ agglomerative clustering][]{agg}.
However, clustering algorithms could not handle the very wide disparity in size
between the T1LM sample (29,237 quasars) and the T1CERQ sample (72 quasars).
Since T1CERQs are a very small portion ($0.25\%$) of the total quasar sample, clustering
methods were either not able to find T1CERQs as a separate cluster or, if they could,
the uncertainties in the obtained labels were high.

Instead, we used a Local Outlier Factor (LOF)\footnote{We used the LOF implementation in
\href{https://scikit-learn.org/stable/modules/generated/sklearn.neighbors.LocalOutlierFactor.html}{\texttt{sklearn}} \citep{scikit}.}
analysis \citep{lof-paper} which quantifies the level of
distinctness in T1CERQs. The LOF has had other uses in astronomy: for example,
detecting unusual spectra in SDSS \citep{wei-2013} and
distinguishing supernovae candidates from massive galaxies \citep{tu-2010}.
LOF measures the extent to which
a data point is isolated with respect to its neighbours
by comparing the  local reachability density of an object  to
the local reachability density  of its k-nearest neighbours using the following score:
\begin{equation}\label{eq:lof}
 LOF_k(A)  =  \frac{1}{\rho_k(A)}\frac{\sum_{B\in \mathbb{N}_k(A)} \rho_k(B)}{\|\mathbb{N}_k(A)\|}.
\end{equation}
This is otherwise called the LOF score for the k-nearest neighbours of point $A$.
$\rho_k(A)$ (or $\rho_k(B)$) is the local reachability density of the k-nearest
neighbours of $A$ (or $B$), defined by:
\begin{equation}\label{eq:rho_k}
  \rho_k(A) =  \frac{\|\mathbb{N}_k(A)\|}{\sum_{B\in \mathbb{N}_k(A)} RD_k(A,B)},
\end{equation}
where $\mathbb{N}_k(A)$ (or $\mathbb{N}_k(B)$) is the set of all k-nearest neighbours of the point $A$ (or $B$).
$\|\mathbb{N}_k(A)\|$ is the number of objects in $\mathbb{N}_k(A)$. $RD_k(A,B)$ is the
reachability distance between point A and B defined by:
\begin{equation}\label{eq:RD}
  RD_k(A,B) = \max\{\mathbb{D}_k(B),\ d(A,B)\}.
\end{equation}
$d(A,B)$ in Eq.~\ref{eq:RD} is the Euclidean distance between point $A$ and $B$ in the normalised 2 or 3D space.
$\mathbb{D}_k(B)$ is the set of
all distances between point $B$ and $\mathbb{N}_k(B)$. \par

 For example,
the density around a  data point, deep in a dense cluster of points,
is very similar to the density of its neighbourhood; this results in LOF$\sim1$.
If the data point is located somewhere
denser than its nearest neighbours, then it has LOF$<1$.
A point where the average density of the neighbours is higher than that of the point has
LOF$>1$, corresponding to the expected behaviour for a small cluster separate from the main group.

The LOF is defined as a function of the number of nearest
neighbours, $k$, which sets the scale or resolution of the cluster
searched for. Thus translated into our analysis, $k$ provides
information about the size of the putative T1CERQ cluster.

\subsubsection{Mock Data Analysis in 2D}
\begin{figure}
  \centering
  \begin{subfigure}{0.5\textwidth}
  \includegraphics[width=\linewidth]{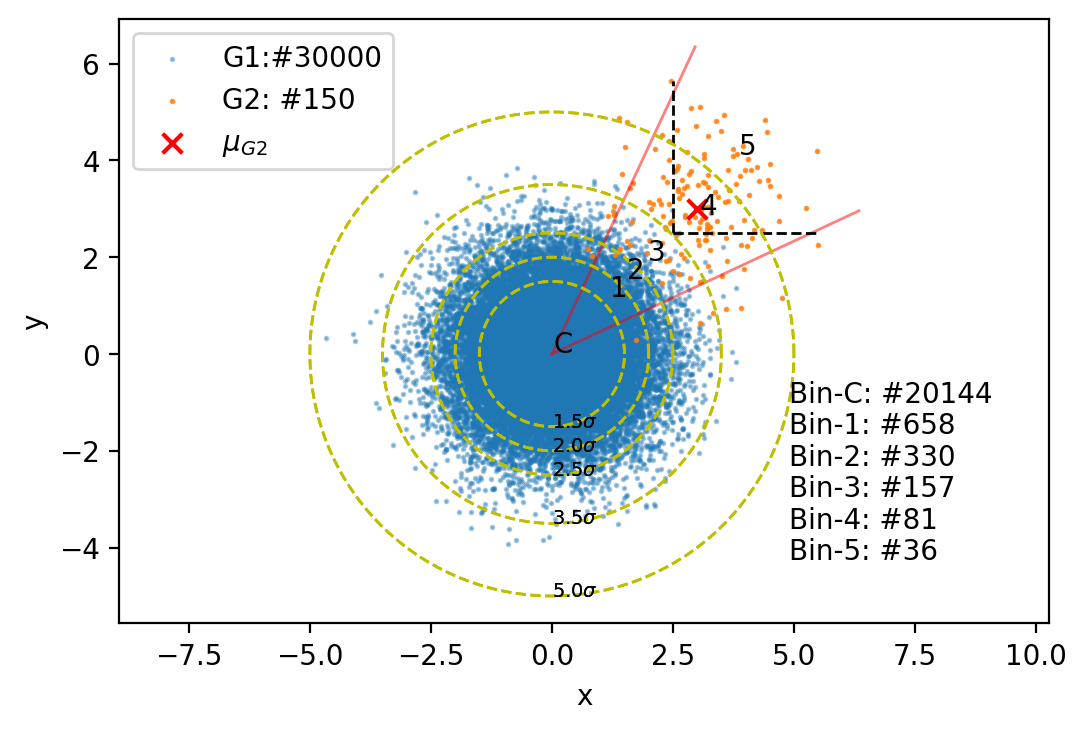}

  \caption{}\label{fig:2d-mock-bin}
  \end{subfigure}
  \begin{subfigure}{0.5\textwidth}
  \includegraphics[width= \linewidth]{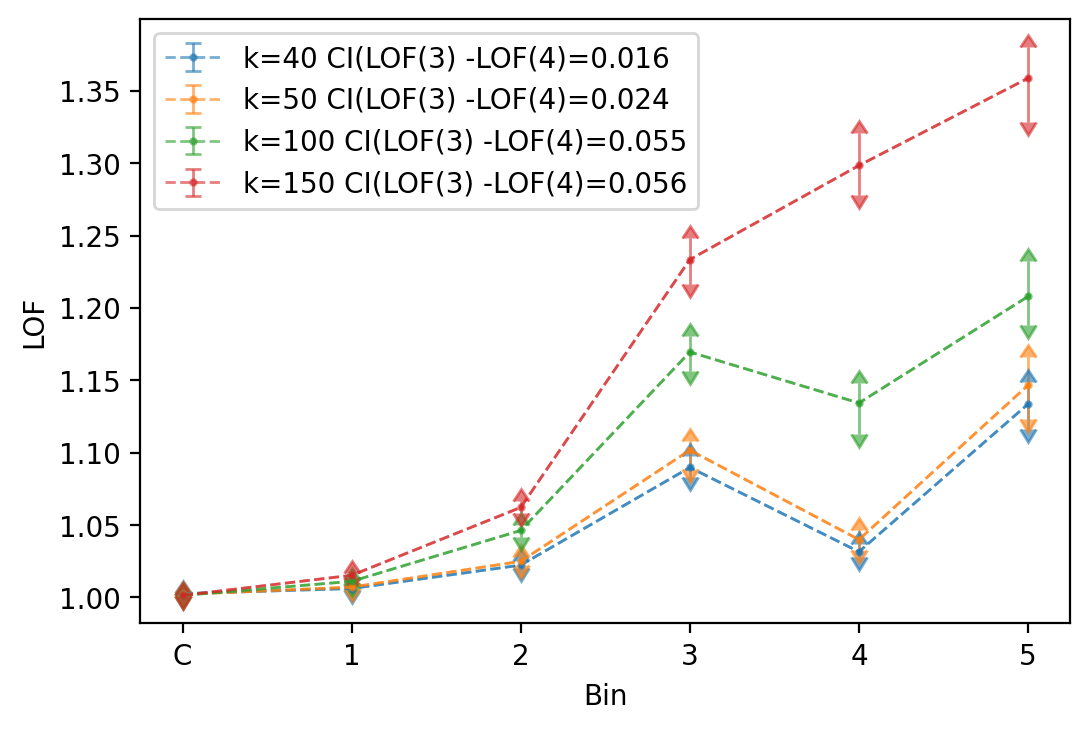}
  \caption{}\label{fig:mock-lof-2d}
  \end{subfigure}
  \caption{Top (a): 2D density binning on a mock data set composed of two
   Gaussian populations: $G_1$: $\mathcal{N}(\mu=[0,0], \sigma=[1,0;0,1])$ with $30000$
   points (blue dots) and $G_2$: $\mathcal{N}(\mu=[3,3], \sigma=[1,0;0,1])$ 
   with $100$ points (orange dots). The red cross shows the center of $G_2$ sample. 
   The yellow circles are at constant distances from the center of $G_1$ of $1.5\sigma$, $2\sigma$, $2.5\sigma$, $3.5\sigma$, and $5\sigma$.
   The data points inside the dashed lines ($x,y>2.5$) are those which would
    be selected as mock T1CERQs following the procedure outlined in \protect\S\ref{sec:wedge-cone}.
   Bottom (b): Median LOF scores and their uncertainties in the bins shown in the top
   panel for nearest neighbours of $k=40, 50, 100, 150$.}
  \label{lof2dmock}
\end{figure}

To better illustrate the behaviour of the LOF$(k)$ score on known distributions of
data points and for different $k$-nearest neighbours, we created 100 mock 2D data
sets by making  100 draws from two overlapping Gaussian distributions, $G_1$ and $G_2$.
To make each mock data set we draw 30000 data points from
 ($G_1$: $\mathcal{N}(\mu=[0,0], \sigma=[1,0;0,1]$)
and 200 data points from ($G_2$: $\mathcal{N}(\mu=[3,3], \sigma=[1,0;0,1]$)),
which in total gives us 100 mock data sets consisting of 30200 data points each.
We chose the same covariance matrix for $G_1$ and $G_2$ for
simplicity. However, the distance between the centers of $G_1$ and $G_2$ imitates
the distance between the median of $i-W3$ in T1LM and the median of
$i-W3$ in T1CERQs.
Similar to the $i-W3\ge4.6$ and REW(\civ)$\ge100$\AA \ cuts which define T1CERQs, we define a core $G_2$
sample (c$G_2$) with $x,y>2.5\sigma$. The average population of
c$G_2$ among our 100 mock data sets is 96 (see Figure \ref{fig:2d-mock-bin}). Moreover,
on average only 1 data point from $G_1$ belongs to c$G_2$, while
on average 104 data points from $G_2$ lay outside of the $x,y>2.5\sigma$ cuts,
showing the level of blending between the $G_2 $ and $G_1$ populations.

We create a wedge, following the same procedure as \S\ref{sec:wedge-cone},
for our mock data.
We are interested in the behaviour of data
in a wedge directed from the median of the bigger population
($G_1$ in the mock 2D data, T1LM in the real dataset)
towards the smaller population (c$G_2$ in the mock 2D data, T1CERQs in the real dataset).
 Note that the centre of $G_2$ is $3\sigma$ away from the centre of $G_1$.
 The opening angle
 for the mock wedge (see $\S$\ref{sec:wedge-cone}) is 20$^{\degr}$
 to be close to the opening angle in 2D of the real data (18.5$^{\degr}$)
 The corresponding unit vector that is directed from the center of $G_1$ to
 the center of $G_2$ is $\hat{v}_{cG_2}=(1/\sqrt{2}, 1/\sqrt{2})$.
To see how LOF$_k$ changes  along  $\hat{v}_{cG_2}$, we binned the wedge as a 
function of distance from the center of $G_1$, with bins shown in Figure \ref{fig:2d-mock-bin}.

We then calculated median LOF scores in each bin of
the mock 2D data set using nearest neighbours (ie, cluster size)
$k=40, 50, 100$, and $150$ for each of our 100 mock data sets.
We plotted the corresponding 68\% confidence intervals
for median LOF scores within each bin in Figure \ref{fig:2d-mock-bin}.
The LOF scores for the mocks have a local minimum
in bin 4, well beyond the 68\% confidence intervals of bin 3 and bin 5 and also
consistent with the confidence interval of the difference between LOF scores of bin 3
and bin 4 (CI(LOF(3)) - CI(LOF(4)) when $k=40$ or $k=50$, but not when $k=100$ or $150$.

This local minimum in LOF score is caused by
the local over-density in bin 4 from $cG_2$. A point in bin 6 close to the centre of
$G_2$ is located in a denser region compared to the average point located in
bin 3, where the transition between $G_1$ and $G_2$ happens; thus the average
LOF score in bin 4 is smaller than in bin 3. In the terminology of the literature, bin 4
includes \emph{locally less outlier} data points. Data points in bin 5 are far
from the centre of $G_2$; as a result, data points in bin 4
are on average also \emph{locally less outlier} than data points in bin 5. These two
observations explains the dip in the LOF score of bin 4 for $k=40$ and $50$.

However, the local over-density in bin 4 (i.e. local minimum in median LOF score)
 is less significant when we consider more neighbours (i.e $k=100$).
 This is because the population size of $cG_2$ in the sample is $96$ points.
A larger cluster includes much of $G_1$ in addition to $G_2$ (see Eq. \ref{eq:lof}).
Thus LOF$_{100}$ and LOF$_{150}$, by incorporating more nearest neighbours for a data
 point in bin 4, do not show a local minimum in the median LOF score.
 A significant local minimum in the LOF scores in a specific region of parameter
space can therefore be used to find the boundary between two populations,
even though they have dramatically different sizes.

We confirmed that this local minimum did not occur in other mock datasets
without two clearly separated populations. We tested the LOF score variation in a
single Gaussian population ($G_1$) with a normal distribution
of $\mathcal{N}(\mu=[0,0], \sigma=[3,0;0,3]$).
We generated LOF$_k$ scores for $k = 40, 50, 100$ and $150$ in $100$ draws from $G_1$
each with $30,000$ data points,
and used the same binning as for the mock dataset containing two Gaussian distributions.
As before we looked at the median LOF$_k$ score and 68\%  confidence intervals around it.
The corresponding plot to Figure~\ref{fig:mock-lof-2d}
never showed a local minimum in the LOF$_k$ score.

\subsubsection{Mock Data Analysis in 3D}
To confirm that this dip also occurs in a mock 3D dataset, we performed a similar
 analysis for 3D Gaussian distributions $G_1: \mathcal{N}(\mu=[0,0,0], \sigma=[1,0,0;0,1,0;0,0,1])$
with $30000$ points and $G_2 \mathcal{N}(\mu=[3,3,3], \sigma=[1,0,0;0,1,0;0,0,1])$
with $150$ data points. We generated $100$ mock data sets and used the same cuts
 for building $cG_2$ ($x,y,z\ge2.5\sigma$). On average the population of $cG_2$
  in our 100 mock data sets is $50$. For comparison, the population size of 
  T1CERQs with the additional kt$_{80}$(\civ)$\ge 0.33$ cut is $52$. We never 
  have a point from $G_1$ in the $x,y,z\ge2.5\sigma$ region, but on average 100 data points from $G_2$.

We used a simple binning procedure similar to the one used for our mock 2D data. We defined a cone
along $\hat{v}=[1/\sqrt{3}, 1/\sqrt{3}, 1/\sqrt{3}]$, directed from the centre of
$G_1$ to the center of $G_2$ with an opening angle of $20^{\degr}$. We defined
a central bin C where $r\le1\sigma$. Bin 1-5 are  within the cone and
between two spheres as follows: bin 1, $1\sigma\le r \le 1.5\sigma$; bin 2,
$1.5\sigma \le r \le 2.5\sigma$; bin 3, $2.5\sigma\le r \le 4.8\sigma$; 
bin 4, $4.8\sigma\le r \le 7\sigma$; bin 5, $r \ge 7\sigma$.

\begin{figure}
\includegraphics[width=\linewidth]{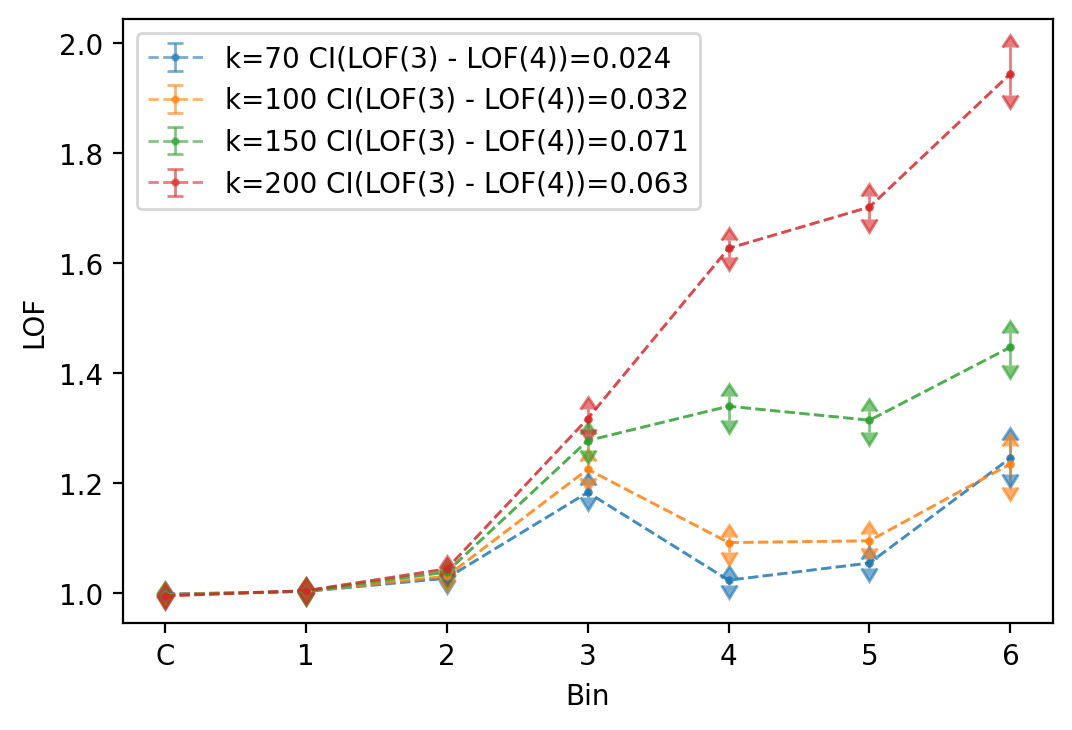}
\caption{Mock 3D with 2 Gaussian population of $G_1: \mathcal{N}(\mu=[0,0,0], \sigma=[1,0,0;0,1,0;0,0,1])$
with $30000$ points and $G_2 \mathcal{N}(\mu=[3,3,3], \sigma=[1,0,0;0,1,0;0,0,1])$
with $150$ data points for the number of nearest neighbours: $k=70, 100, 150$, and $200$.
CI(LOF(3)-LOF(4)) in the legend refers to the the 68\% confidence interval for the
difference of LOF score between bin 3 and bin 4 for each $k$. }
\label{fig:mock-3d}
\end{figure}

Figure \ref{fig:mock-3d} shows the median LOF score in each bin for different
nearest neighbours of $k=70, 100, 150$, and $200$. The 68\% confidence intervals
for each median LOF score is shown as the error bar. The decrease in the LOF
score from bin 3 to bin 4 is more than the 68\% confidence intervals of each bin
and also more than the 68\% confidence interval of the difference between
the average LOF score in bin 3 and bin 4. As a result the
local dip in the LOF score of bin 6 is significant to at least the 68\% level.

Having demonstrated that a signature of two mixed populations (in 2D and 3D) is
a dip in the LOF$_k$ computed along the vector towards the smaller
 population, at a nearest neighbour value smaller than the size of the smaller
 population, we continue to analyse our real quasar sample.

\section{Results}
\label{sec:results}

\subsection{Density in the 2D parameter space}\label{sec:density}
\begin{figure}
    \centering
    \includegraphics[width= \linewidth]{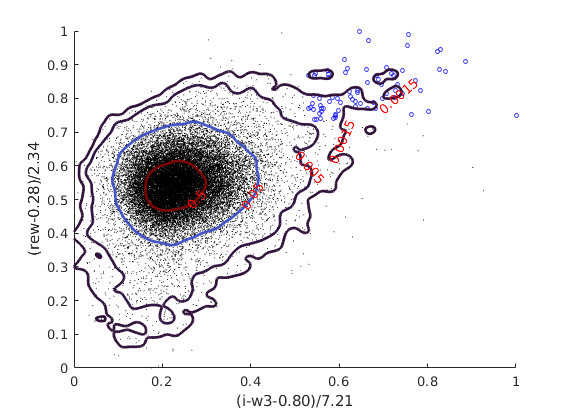}
    \caption{Density of quasars in a normalised $i-W3$, REW(\civ) space. Density contours are shown relative to the maximum density at
    $\rho/\rho_{max}=0.5, 0.05, 0.005, 0.0015$. Blue circles are T1CERQs.
    Black dots are the other T1LMs. }
    \label{density-2d}
\end{figure}

 Our first objective is to gather evidence to determine whether T1CERQs are
  part of a separate population or extreme examples of T1 quasars,  probing 
  the tail of the main distribution. As a first, simple attempt to answer
  this question, Figure \ref{density-2d} visualises the quasars in the 2D
  parameter space of $i-W3$ colour, REW(\civ), normalised as explained
  in \S \ref{sec:wedge-cone}. It is visually apparent that the T1CERQs
  are over-dense compared to other regions of parameter space at a
  similar distance from the main quasar locus.  To quantify how much, we
  computed the density of quasars in parameter space using kernel density
  estimation (KDE) with a Gaussian kernel. We want to compare the density
  of the parameter space to the high density region near the median of
  T1LM sample, and so we plotted density contours relative to the
  maximum density. For the KDE smoothing bandwidth we applied 
  Silverman's rule of thumb to obtain the bandwidth for each dimension separately:
 \begin{equation}\label{band}
   h_i = \sigma_i \Bigg(\frac{4}{N(d+2)}\Bigg)^\frac{1}{d+4}.
 \end{equation}
 Here, $\sigma_i$ is the standard deviation for the \emph{i}-th dimension of our normalised parameter space, $N$ is the number of objects (29237 for the T1LM sample), and $d$ is the number of dimensions; 2 in 2D parameter space and 3 in 3D parameter space.
Given $\sigma_{i-W3}=0.077$ and $\sigma_{REW(\mathrm{CIV})}=0.084$, we obtained
$h_{i-W3}^{2D}=0.014$ and $h_{REW(\mathrm{CIV})}^{2D}=0.015$.

\begin{figure}
    \centering
    \includegraphics[width= \linewidth]{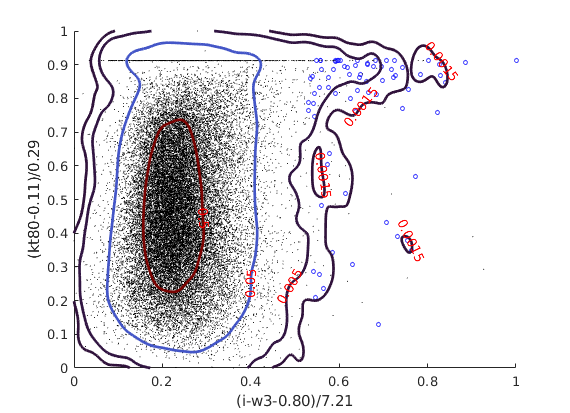}
    \caption{Density of quasars in a normalised $i-W3$, kt$_{80}$ space.
    Density contours are shown relative to the maximum density at
    $\rho/\rho_{max}=0.5, 0.05, 0.005, 0.0015$. Blue circles are T1CERQs.
    Black dots are the other T1LMs. }
    \label{density-2dkt80}
\end{figure}

The density contours with $\rho > 0.05 \rho_{max}$ in Figure \ref{density-2d}
 are similar in shape, and show the shape of the
  main quasar  locus. However, the contours at lower densities are elongated in the
   direction of the T1CERQs (blue circles in
    Figure \ref{density-2d}), including some
     mild local density maxima caused by T1CERQs. 
     The low number of samples in this region means that the
      density contours are somewhat noisy, but
it is apparent that the T1CERQs are an over-density in this
 parameter space. Figure~\ref{density-2dkt80} shows a similar
  density trend in the $i-W3$, kt$_{80}$ plane.
Here the over-density near the T1CERQs is even 
more apparent: the lowest density contour is extended
at kt$_{80} \sim 0.35$ towards high $i-W3$.

A possible explanation for these outer contours is the non-linear effect of dust reddening on the colour distribution of quasars \citep{richards-2001}.
However, Fig. 11 of H17 shows that the typical SED of ERQs is very different from the SED of dust-reddened quasars without the strong \civ~line characteristic of core ERQs.
Core ERQs have SEDs which are much flatter
in the rest-frame UV than suggested by their red $i-W3$ colours, while non-core ERQs exhibit a sharp decline in the near UV with only moderately red colours across the near IR, similar to type 1 QSOs reddened by dust extinction (H17).

%
%
%
%

\subsection{Median Spectra}\label{sec:med-spec}

\begin{figure}
\centering
    \includegraphics[width=\linewidth]{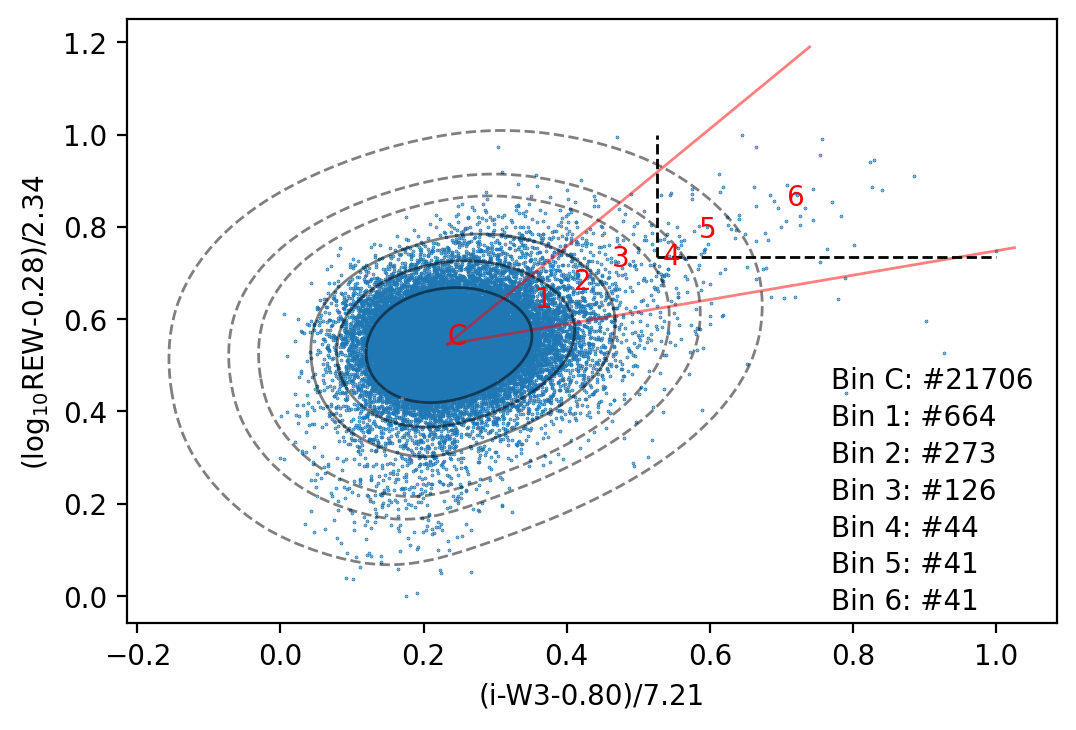}
\caption{A binned wedge along $\vec{v}_{T1CERQ}$ towards the T1CERQ sample, with bins defined by density contours. The population of each bin is provided. Bin-C is enclosed by the innermost solid line contour at $0.3\rho_{max}$. 2nd and 3rd contours are at the levels of 0.03 and 0.01 of $\rho_{max}$. The three outer dashed line contours are
$\times$1.35, $\times$1.55, and $\times$1.95 enlarged version of the biggest solid line contour.}
\label{fig:bin-wdg-1}
\end{figure}

\begin{figure*}
\includegraphics[width= \linewidth]{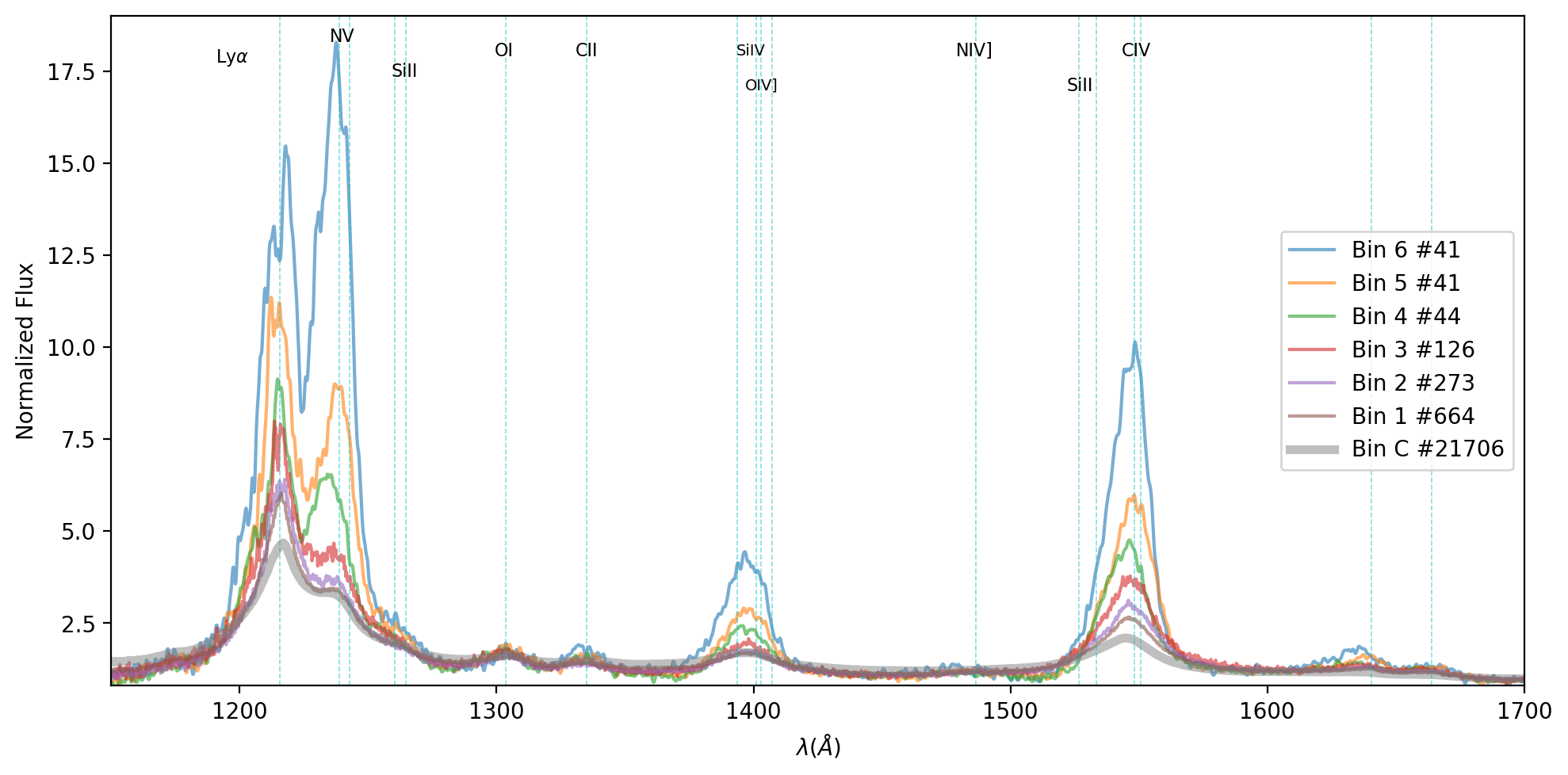}
\caption{Median spectra for bins in the direction of $\vec{v}_{T1CERQ}$.
The bin number and the number of quasars in each bin are shown. As a reminder, T1CERQs are found in bins 5 and 6 and part of bin 4.}\label{fig:2d-medspec-w1}
\end{figure*}

\begin{table*}
\caption{List of median physical properties in each bin from Figure \ref{fig:bin-wdg-1}.
"C" in the 2nd column denotes the central bin.
Column No. shows the number of quasars in each bin. $f_{BAL}$ is the fraction of quasars which contain a visually verified
BAL feature near the \civ \ line. Other columns show the median
values and their standard deviations in each bin.
}

\label{tab:2d-properties}
\begin{tabular}{|l|l|l|l|l|l|l|l|l|l|}
  \hline
    \multicolumn{1}{|c|}{Bin} &
    \multicolumn{1}{c|}{No.} &
    \multicolumn{1}{c|}{$i-W3$} &
    \multicolumn{1}{c|}{REW(\civ)} &
    \multicolumn{1}{c|}{FWHM(\civ)} &
    \multicolumn{1}{c|}{kt$_{80}$(\civ)} &
    \multicolumn{1}{c|}{\fnc} &
    \multicolumn{1}{c|}{$f_{BAL}$} &
    \multicolumn{1}{c|}{Luminosity}  \\ \hline
    C & 21706 & 2.45 $\pm$ 0.35 & 35 $\pm$ 10 & 5400 $\pm$ 1500 & 0.25 $\pm$ 0.05 & 1.54 $\pm$ 0.55 & 0.14 &  46.83 $\pm$ 0.21\\ \hline
    1 & 664 & 3.37 $\pm$ 0.15 & 55 $\pm$ 8 & 4700 $\pm$ 1700 & 0.24 $\pm$ 0.06 & 1.04 $\pm$ 0.46 & 0.22  &  46.76 $\pm$ 0.19\\
    2 & 273 & 3.76 $\pm$ 0.18 & 68 $\pm$ 13 & 4300 $\pm$ 1700 & 0.23 $\pm$ 0.07 & 0.95 $\pm$ 0.55 & 0.29 &  46.79 $\pm$ 0.22\\
    3 & 126 & 4.14 $\pm$ 0.24 & 89 $\pm$ 22 & 3900 $\pm$ 1500 & 0.28 $\pm$ 0.07 & 0.99 $\pm$ 0.60 & 0.32 &  46.86 $\pm$ 0.24\\
    4 & 44 & 4.65 $\pm$ 0.17 & 92 $\pm$ 30 & 3500 $\pm$ 1400 & 0.35 $\pm$ 0.07 & 1.66 $\pm$ 0.78 & 0.52 &  47.02 $\pm$ 0.28\\
    5 & 41 & 5.02 $\pm$ 0.26 & 125 $\pm$ 42 & 3300 $\pm$ 1200 & 0.35 $\pm$ 0.06 & 1.28 $\pm$ 0.76 & 0.29 & 47.09 $\pm$ 0.32\\
    6 & 41 & 5.90 $\pm$ 0.57 & 181 $\pm$ 80 & 3100 $\pm$ 900 & 0.36 $\pm$ 0.06 & 1.74 $\pm$ 0.58 & 0.12 &  47.30 $\pm$ 0.29\\ \hline
  \hline
\end{tabular}
\end{table*}

A second intuitive way to examine exotic line properties is to
make median spectra for the T1CERQ sample. Since we have,
in \S\ref{sec:wedge-cone}, constructed vectors towards the T1CERQ sample,
together with geometric cones which contain the T1CERQ quasars.
We are now able to bin the cones and make median spectra within these bins
and examine how the spectra of T1CERQs change in 2- and 3D parameter space.
Median spectra are made by stacking, after the following pre-processing steps:
\begin{enumerate}[(1)]
	\item Shift the observed flux into the quasar's rest frame.
	\item Normalise the spectrum by the median flux between 1680\AA \ and 1730\AA \ in the rest frame. This region was chosen as the quasar spectrum is mostly free from significant line features.
	\item Interpolate all fluxes onto a logarithmic grid defined between 800\AA \ and 3000\AA.
\end{enumerate}


Since visualisation is easier in 2D, we perform our first median spectrum analysis by
 binning in the normalised parameter space of $i-W3$ and log$_{10}$REW(\civ)
  along the $\vec{v}_{T1CERQ}$ direction described in \S\ref{sec:wedge-cone}.
Figure~\ref{fig:bin-wdg-1} shows the 2D wedge towards $\vec{v}_{T1CERQ}$,
together with the density contours around which we define the bins for our median spectra.
The bin boundaries are chosen to bring
out specific features of the median spectra.  Three inner density contours
at $0.3\rho_{max}$, $0.1\rho_{max}$, $0.03\rho_{max}$ show the shape of the
core of the T1LM sample. The three outermost contours scale the contour
at $0.03\rho_{max}$ by 1.35, 1.55, and 1.95, as the number of quasars
this far from the main locus is too low to accurately estimate density.
The choice of the 1st (1.35) and 2nd (1.55) scale factors gives a bin
that covers the boundary of T1CERQs suggested by H17 (the lower left corner of
REW(\civ)>100\AA \ and $i-W3\ge$4.6 box in Figure \ref{fig:bin-wdg-1}).
The 3rd scale factor (1.95) is chosen so that bins 5 and 6 are equally populated.

Figure~\ref{fig:2d-medspec-w1} and Table \ref{tab:2d-properties} show
that there is an evolution in the line properties of quasars along
$\vec{v}_{T1CERQ}$. The median \civ \ emission line in bins 1
through 3 is symmetric, close to the shape of the median spectrum
 in bin C, the main T1LM quasar locus (shown by the thick grey
  spectrum in Figure~\ref{fig:2d-medspec-w1}).

Table \ref{tab:2d-properties} shows that there is a jump in the \civ~line kurtosis in bin 4: median kt$_{80}$(\civ) is $0.28$ in bin 3
and $0.35$ in bins 4-6. Bin 4 also has an unusually large BAL fraction ($0.52$).
To ensure that these properties are due to the T1CERQ vector and not a function of distance from the quasar locus, we also checked the line properties along $-\vec{v}_{T1CERQ}$ and confirmed that kt$_{80}$(\civ) remained similar to bin C, while the BAL fraction dropped to $0.07$. We confirmed these trends by making median spectra along vectors both clockwise and anti-clockwise of $\vec{v}_{T1CERQ}$, again finding that the line kurtosis remained low and confirming that the $\vec{v}_{T1CERQ}$ direction is unique.

There is also a relatively large jump in \fnc \ from $\sim 1 $ in bins $1-3$ to $1.66$ in bin 4. We found that \fnc \ also increased along $-\vec{v}_{T1CERQ}$, perhaps indicating that this is not intrinsic to T1CERQs, but in this direction the \ion{N}{V} line is weak and the fit is likely to suffer severely from blending with the Lyman-$\alpha$ line.

\subsubsection{3D parameter space }\label{sec:med-spec-3d}

\begin{figure*}
  \centering
  \begin{subfigure}{1.0\textwidth}
	\includegraphics[width=0.5\linewidth]{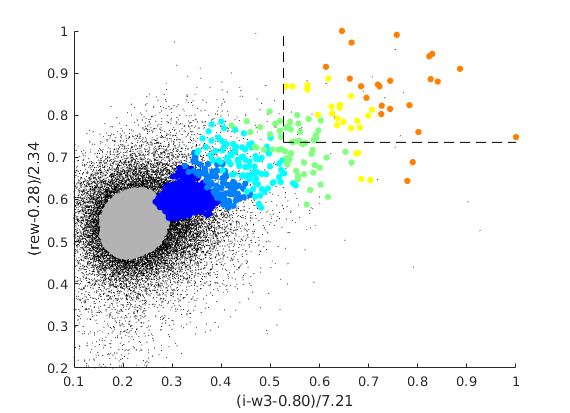}
		\includegraphics[width=0.5\linewidth]{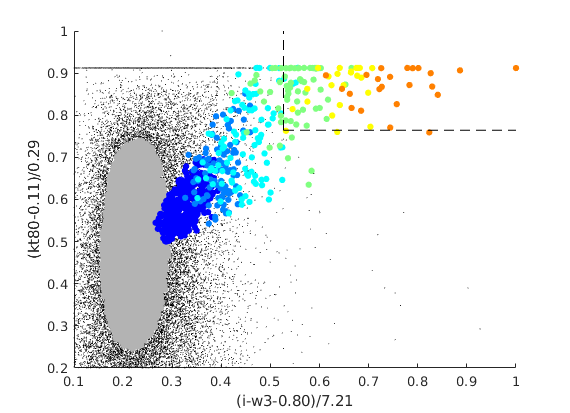}
    \caption{}
  \label{fig:3d-bin}
  \end{subfigure}
\begin{subfigure}{\textwidth}
  \includegraphics[width=\textwidth]{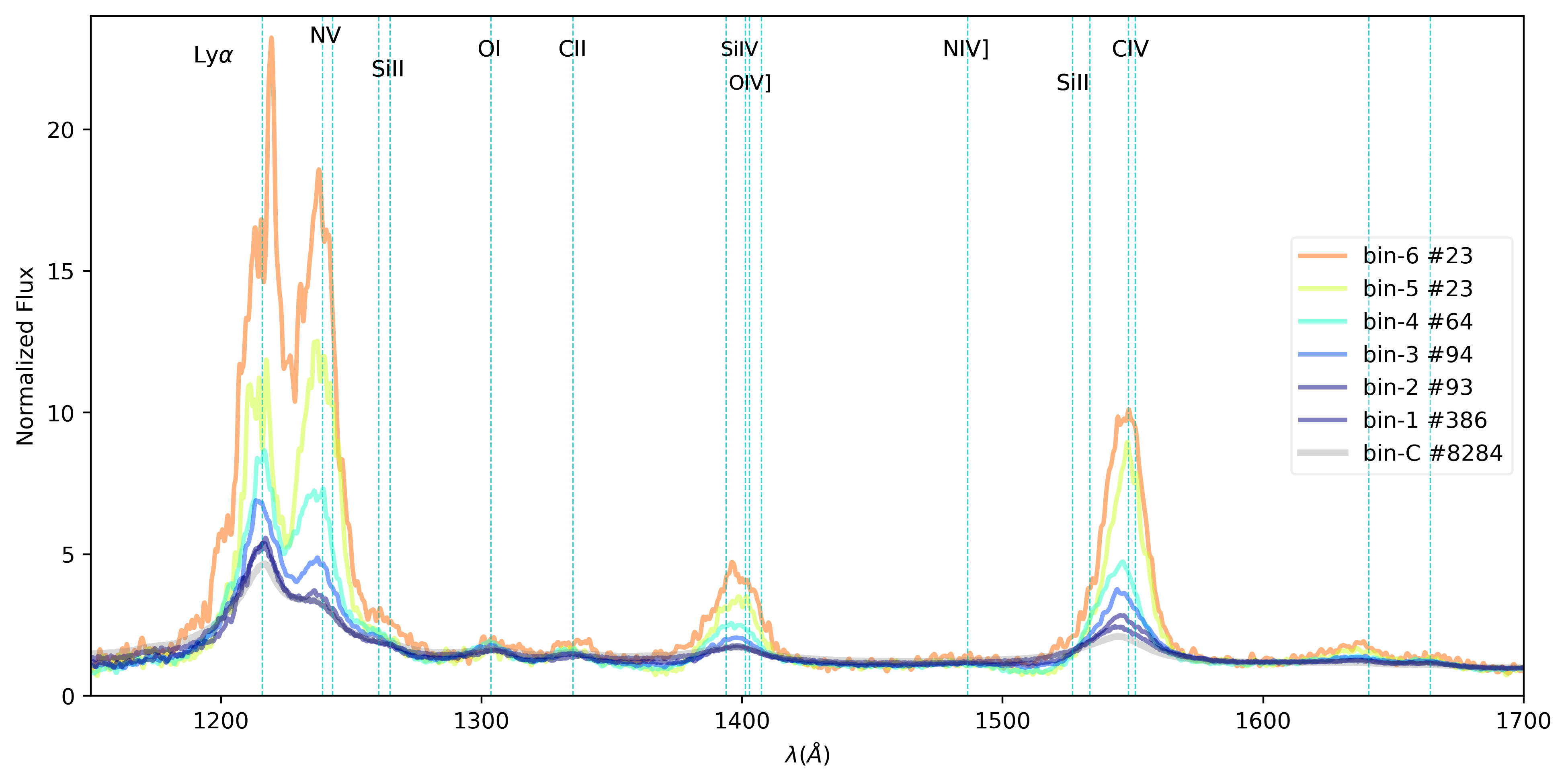}
\caption{}
\label{fig:3d-medspec}
\end{subfigure}
\caption{Top (a): 3D bins along a cone around $\vec{v}_{T1CERQ}^{3D}$. 
The central bin is shown by grey points at the centre. Each bin, separated by 
density iso-surfaces as described in \protect\S\ref{sec:med-spec-3d}, is painted a different colours.
Dashed lines shows the region of $i-W3\ge 4.6$, $\log_{10}$REW(\civ)$\ge 2$, and
kt$_{80}$(\civ)$\ge0.33$ in the min-max normalised space.
Bottom (b): Median spectra for the corresponding coloured objects in each 
bin of the top panel. Spectra colours for each bin match those in Figure~\protect\ref{fig:3d-bin}.}
\end{figure*}

Motivated by the success of our 2D analysis, we made median
spectra in the 3D normalised parameter space
 of $i-W3$, log$_{10}$REW(\civ), and kt$_{80}$(\civ).
The median spectra bins were made within the 3D cone
defined in \S\ref{sec:wedge-cone}. The central bins were again defined by
density iso-surfaces relative to the maximum density and computed
using a KDE as in \S\ref{sec:density}.
The central quasar locus, bin C, was $\rho > 0.5\rho{max}$.
Bin 1 has  $\rho  = 0.5\rho{max} - 0.05\rho{max}$ and bin 2
is $\rho  = 0.05\rho{max} - 0.01\rho{max}$.
As for our 2D binning, we did not use the density iso-surfaces
at lower density levels, because the low numbers of spectra in
these bins make local density estimates too noisy. Instead, we
uniformly enlarged the iso-surface of $0.01\rho_{max}$ by
factors of $1.5$, $2.1$, and $2.5$ to make three extra
surfaces and used these enlarged surfaces for bins 4 though 6.
The expansion factors of $1.5$ and $2.1$ were chosen so that
bin 4 covers H17's boundary for ERQs (the $i-W3\ge4.6$
plane in Figure \ref{fig:3d-bin}). The scale factor
of $2.5$ was chosen so that the last two bins had an equal sized population
(23 quasars in each).

All 3D bins are colour
coded in Figure \ref{fig:3d-bin}. The transparent box
in Figure \ref{fig:3d-bin} shows $i-W3\ge4.6$, $\log_{10}(REW(\civ))\ge100$\AA, and kt$_{80}$(\civ)$\ge0.33$.
Median spectra of these 3D bins are plotted
in Figure \ref{fig:3d-medspec} and Table \ref{tab-3d-properties} summarises the
median physical properties in each bin of Figure \ref{fig:3d-bin}.
As for the 2D analysis, the kurtosis increments in each bin from $1$ to $3$ and saturates at $0.33$ in bin 4, suggesting bin 4 is a good candidate for a boundary
separating a population of red quasars from the main T1LM sample.
\fnc \ is larger in bin $4-6$ compared to bin $1$ though
bin $3$. It is again large in bin C, but this may again be due to blending with Lyman-$\alpha$. As in 2D, bins 3 and 4 have a high fraction of BALs, although this is not true of bin 6. In general the trends in 3D are similar to those in 2D: this extra parameter, however, will be useful in the next sections.

\begin{table*}
  \caption{List of median physical properties in the bins of
  Figure~\ref{fig:3d-bin}. Columns are named as in Table
  \ref{tab:2d-properties}.}
  \label{tab-3d-properties}
  \begin{tabular}{|l|l|l|l|l|l|l|l|l|l|}
    \hline
      \multicolumn{1}{|c|}{Bin} &
      \multicolumn{1}{c|}{No.} &
      \multicolumn{1}{c|}{i-W3} &
      \multicolumn{1}{c|}{REW(\civ)} &
      \multicolumn{1}{c|}{FWHM(\civ)} &
      \multicolumn{1}{c|}{kt$_{80}$(\civ)} &
      \multicolumn{1}{c|}{\fnc} &
      \multicolumn{1}{c|}{$f_{BAL}$} &
      \multicolumn{1}{c|}{Luminosity} \\ \hline
      C & 8284 & 2.40 $\pm$ 0.21 & 35 $\pm$ 6 & 5600 $\pm$ 1300 & 0.25 $\pm$ 0.03 & 1.60 $\pm$ 0.50 & 0.11  &  46.85 $\pm$ 0.21\\
      1 & 386 & 3.10 $\pm$  0.20 & 47 $\pm$ 7 & 5400 $\pm$ 1400 & 0.28 $\pm$ 0.01 & 1.22 $\pm$ 0.44 & 0.16 &  46.79 $\pm$ 0.18\\
      2 & 93 & 3.69 $\pm$  0.21 & 58 $\pm$ 10 & 4500 $\pm$ 1800 & 0.30 $\pm$ 0.02 & 1.09 $\pm$ 0.55 & 0.30 &  46.80 $\pm$ 0.22\\
      3 & 94 & 4.08 $\pm$  0.33 & 78 $\pm$ 18 & 3900 $\pm$ 1600 & 0.32 $\pm$ 0.03 & 1.29 $\pm$ 0.58 & 0.52 &  46.91 $\pm$ 0.22\\
      4 & 64 & 4.74 $\pm$  0.33 & 93 $\pm$ 24 & 3500 $\pm$ 1300 & 0.36 $\pm$ 0.02 & 1.68 $\pm$ 0.72 & 0.55 & 47.19 $\pm$ 0.27\\
      5 & 23 & 5.43 $\pm$  0.35 & 142 $\pm$ 44 & 3000 $\pm$ 1400 & 0.37 $\pm$ 0.01 & 1.78 $\pm$ 0.71 & 0.30 & 47.20 $\pm$ 0.33\\
      6 & 23 & 6.17 $\pm$  0.61 & 209 $\pm$ 92 & 3500 $\pm$ 1100 & 0.36 $\pm$ 0.01 & 1.74 $\pm$ 0.65 & 0.04 & 47.49 $\pm$ 0.27\\
      \hline\end{tabular}
\end{table*}

\subsection{Local Outlier Factor Analysis}\label{sec:lof}

We showed in \S\ref{sec:density}
that T1CERQs are an over-density when compared to other quasars at a
comparable distance from the centre of the main population, and in
\S\ref{sec:med-spec} we found that there was an increase in kurtosis
around the fourth bin from the central T1LM sample.
In this section, we quantify the distinctness of T1CERQs from the main
T1LM sample using LOF, and examine the 2D and 3D candidate boundaries we found in
\S\ref{sec:med-spec}. As a reminder, in \S\ref{sec:mocklof} we showed that a
signature of two distinct populations is a dip in the LOF score.

\subsubsection{LOF Analysis in 2D}
We now proceed to analyse the full T1LM sample with LOF along a vector directed
towards the T1CERQ, $\vec{v}_{T1CERQ}$.
We use the bins depicted in Figure~\ref{fig:bin-wdg-1}, and
compute a median LOF score in the \emph{normalised} 2D space of $i-W3$
and $\log_{10}$REW(\civ)$)$. There is a (small) dip in the median LOF
score for bin 4. This is interestingly consistent with the results from median
spectra in \S\ref{sec:med-spec}, where we saw that bin 4 was also associated with unusual spectral properties. The magnitude of the dip for $k = 40$ is $0.019$, which is somewhat less than the 68\% uncertainty of the LOF score in our 2D mock data analysis ($\sigma($LOF(bin 3)-LOF(bin 4)$)=0.022$). The 2D LOF analysis thus provides indications that the T1CERQs may be a separate population from the main T1LM, but is by no means definitive.

\begin{figure}
  \centering
  \includegraphics[width=\linewidth]{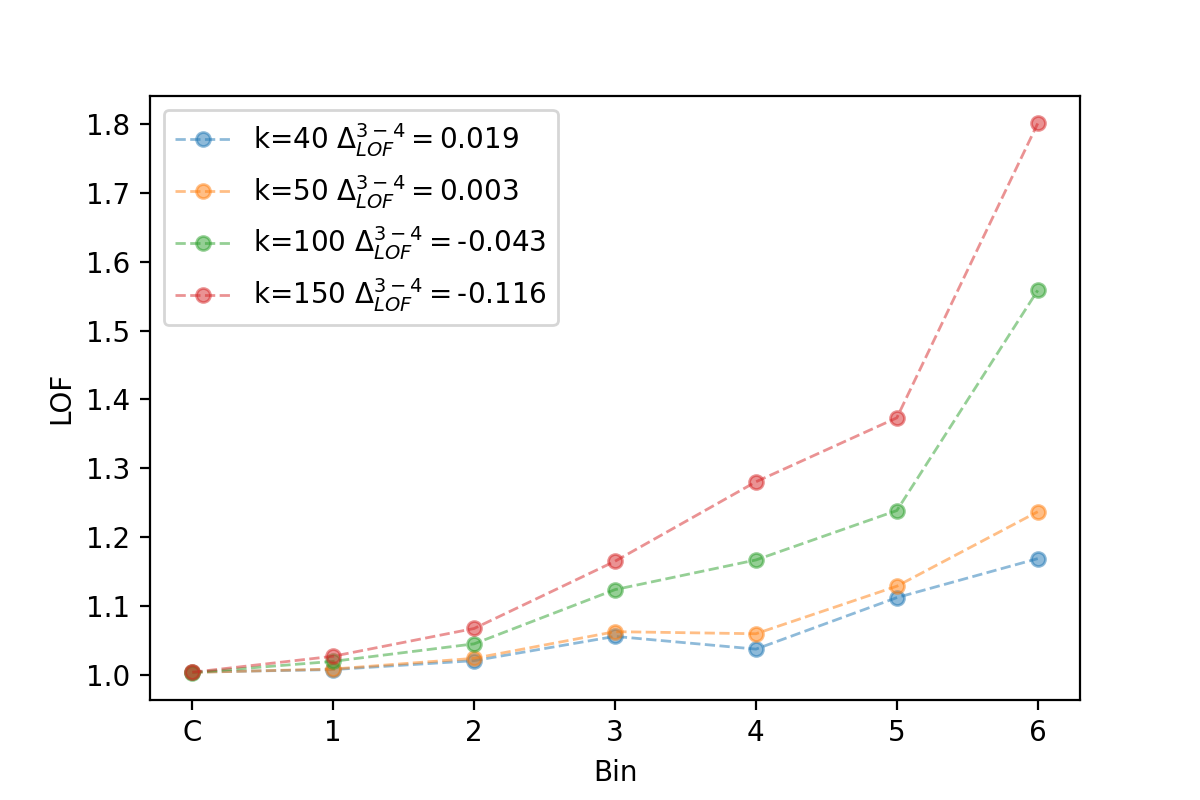}
  \caption{Median LOF score in each bin within the wedge of Figure~\ref{fig:bin-wdg-1}, along the vector $\vec{v}_{T1CERQ}$ between the centroid of T1LM and T1CERQ, for different numbers of nearest neighbours (k).}
  \label{fig:2d-lof}
\end{figure}

Figure~\ref{fig:2d-lof} also shows the dependence of LOF score on neighbour number, $k$. for $k=40, 50, 100$, and $150$. The LOF score falls from bin 3 to bin 4 when the number of nearest neighbours is $40$ or $50$, and monotonically increases for $k=100$ or $150$.
The LOF score thus suggests that a putative separate population of T1CERQs would have a population between $50$ and $100$, in good agreement with H17, who identified a population of $72$ T1CERQs.

\subsubsection{LOF Analysis in 3D}
We performed a similar LOF analysis in 3D, adding kt$_{80}$(\civ)
to $i-W3$, $\log_{10}($REW(\civ)$)$ and using a normalised space.
Figure~\ref{fig:3d-bin} once again shows a dip in the median LOF
score for bin 4. This fall in the LOF score from bin 3 to bin 4 is
$0.123$ for $k=70$, significantly larger than the 68\% uncertainty of the
similar bin in the mock 3D data analysis ($\sigma($LOF(bin 3)-LOF(bin 4)$)=0.025$),
as Figure \ref{fig:mock-3d} shows. The LOF analysis in 3D is thus
evidence that there is a separate population of quasars along the vector to T1CERQs.
Interestingly, Figure~\ref{fig:mock-3d} shows that the dip in the LOF score persists for $k \leq 150$, suggesting that the separate population may be somewhat larger than that found in H17.

\begin{figure}
  \includegraphics[width= \linewidth]{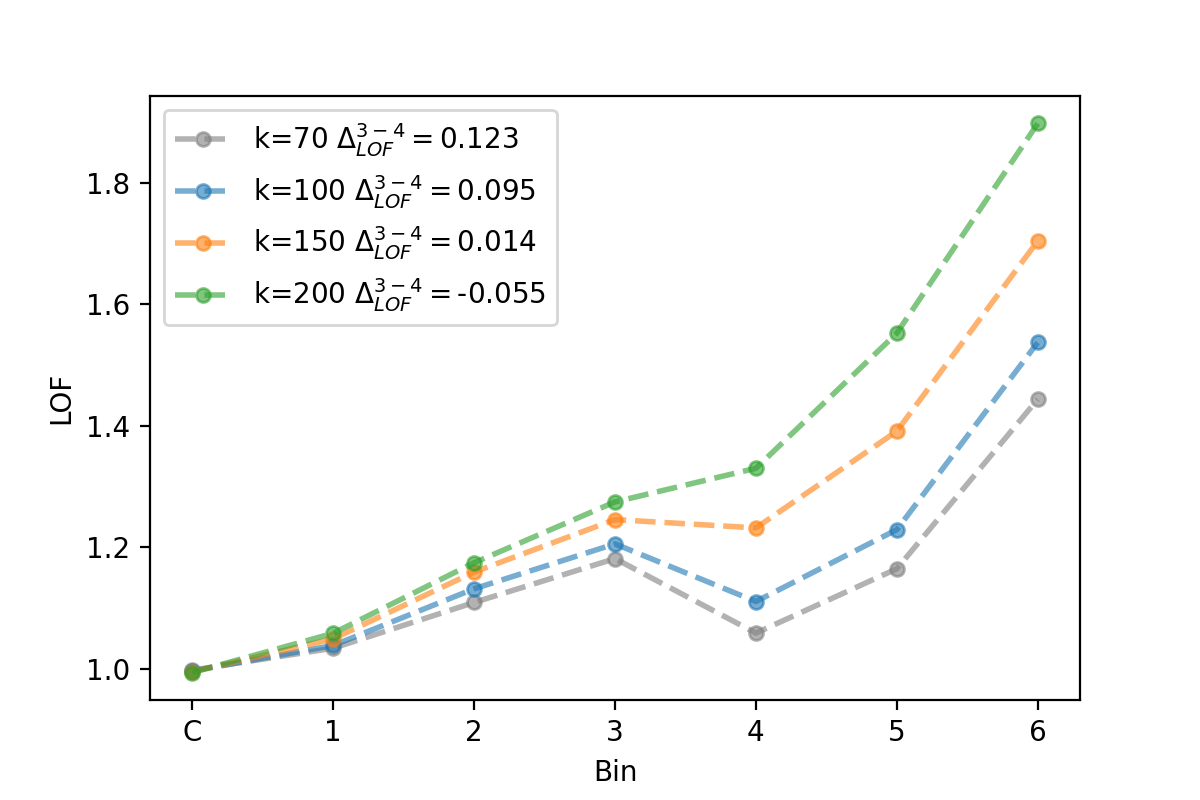}
  \caption{Median LOF score in each 3D bin of Figure~\ref{fig:3d-bin}, along the 3D vector $\vec{v}_{T1CERQ}^{3D}$ between the centroid of T1LM and T1CERQ, for different numbers of nearest neighbours (k).}
  \label{fig:3d-lof}
\end{figure}

\subsection {Selecting T1BERQs in 3D}\label{sec:3d-boundary}

H17 found a sub-population of quasars in 2D space, the T1CERQs.
Our median spectra and LOF analysis has provided quantitative evidence
that this sub-population is distinct from the general trend of
the T1LM sample, especially when viewed in the 3D parameter space
of $i-W3$, $\log_{10}($REW(\civ)$)$ and kt$_{80}$(\civ).
There are also indications in the LOF score that the sub-population is
moderately larger than the T1CERQ set found by H17. In this section we
will design 3D criteria which optimises the selection of these objects.
We call our new subset of quasars Type 1 boxy \civ \ emission line extremely red quasars (T1BERQs).

The choice of our kt$_{80}$(\civ) parameter space is also motivated by
Figure~\ref{matrix-hist}, which suggests that there are a small number of
low kt$_{80}$(\civ) objects within the T1CERQ class and that a minimum
kt$_{80}$ condition will produce a purer sample.
Here we outline our recipe for selecting T1BERQs, summarizing steps
introduced in earlier sections:
\begin{enumerate}[(1)]
\item Normalise the parameter space of ($i-W3$, $\log_{10}$(REW(\civ)), kt$_{80}$(\civ)) with a Min-Max scaler (discussed in \S\ref{sec:wedge-cone}).
\item Define $\vec{v}_{T1CERQ}^{3D}$, a vector
from the median of the T1LM sample to the median of those points satisfying
$i-W3\ge 4.6$, REW(\civ)$\ge 100$\AA, and kt$_{80}$(\civ)$\ge 0.33$), in the normalised space.
\item Find a cone along $\vec{v}_{T1CERQ}^{3D}$, with a tip located at the median point of T1LM, and an opening angle so that the cone includes all quasars satisfying  $i-W3\ge 4.6$, REW(\civ)$\ge 100$\AA, kt$_{80}$(\civ)$\ge 0.33$.
\item Using KDE, find density iso-surfaces and bin the cone in the previous
step by successive iso-surfaces.
One of the bins passes through an initial
guess about the boundary of the desired population (here $i-W3\ge 4.6$,
REW(\civ)$\ge 100$\AA, and kt$_{80}$(\civ)$\ge 0.33$).
\item Calculate the LOF score in each bin.
\item Find the bin showing a local minimum in LOF score.
\item  Repeat steps (4) to (6) varying the inner and outer boundaries of the candidate
 bin found in step (6) and find the bin which shows the largest
 decrease in LOF scores as compared to the neighbour bin located closer to the centre of the T1LM sample.
\item Find a plane perpendicular  to $\vec{v}_{T1CERQ}^{3D}$ and tangent to
the inner boundary of the optimum bin in step (7).
\item Define the boundaries of the T1BERQs using the common region between the
plane of step (8) and the cone of step (3).
\end{enumerate}

\begin{figure*}
  \centering
    \includegraphics[width=0.45\linewidth]{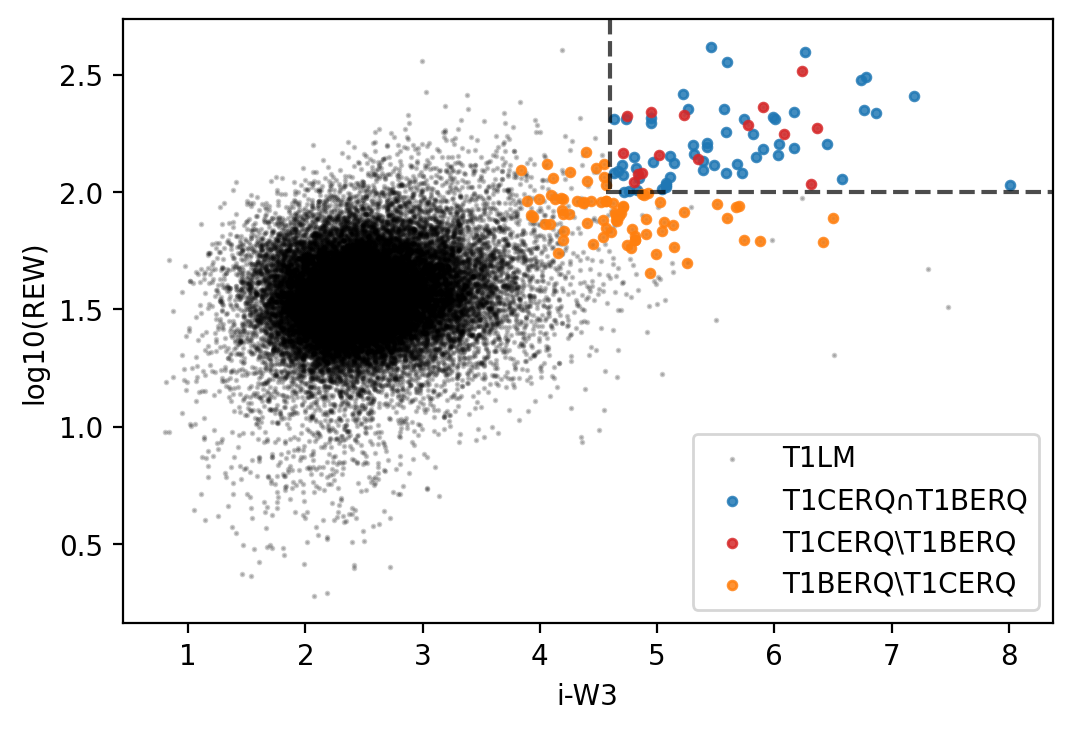}
    \includegraphics[width=0.45\linewidth]{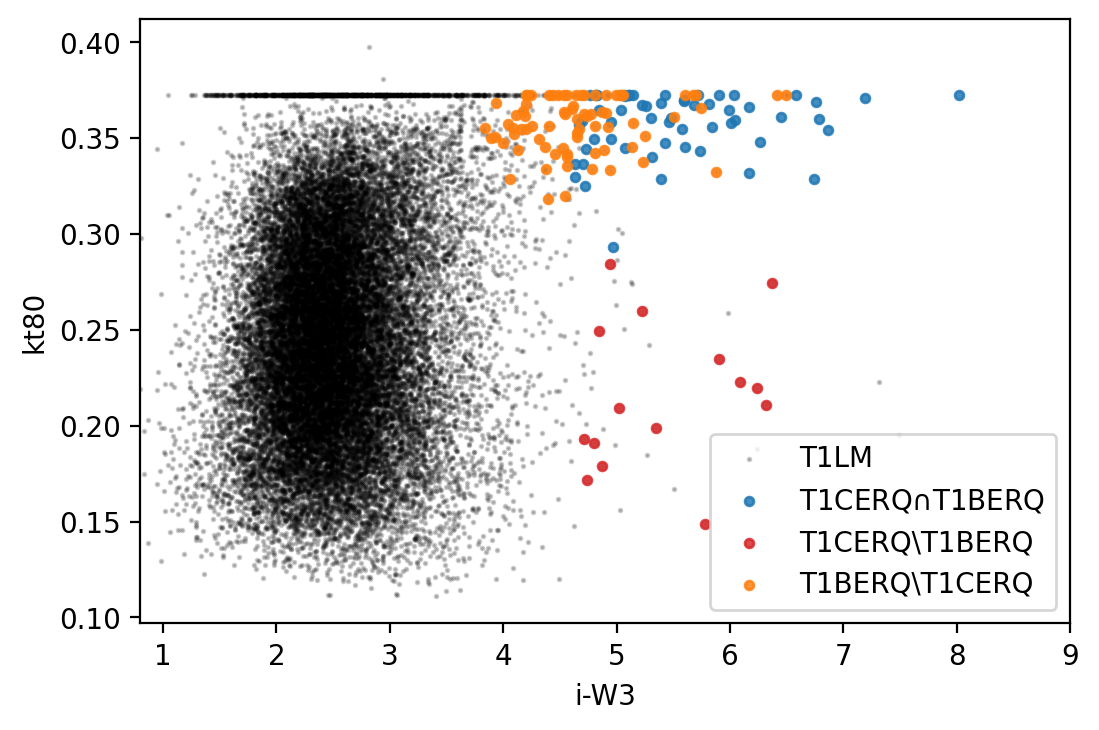}
    \caption{(Left) Projection of the 3D selection of T1BERQs into
    (i-W3, $\log_{10}($REW(\civ)) space. (Right) Projection of the
    3D selection of T1BERQs into (i-W3, kt$_{80}$(\civ)) space.
    Blue dots belong to the intersection of T1CERQs and T1BERQs. Red dots are T1CERQs which are not T1BERQs. Orange dots are T1BERQs  which are not T1CERQs. T1LMs which are not T1CERQS or T1BERQs are shown by black dots.}
    \label{fig:3d-boundary}
\end{figure*}

Following this procedure, we found all quasars in a cone with a tip at (in our normalised 3D parameter space) $(0.23, 0.54, 0.47)$ and an opening angle of $19.6^{\circ}$,
the same cone as in \S\ref{sec:wedge-cone}. The bin with a minimum LOF score was bin 4, as in \S\ref{sec:3d-boundary}, and the optimised, adjusted boundary between bins $4$ and $3$ was expanded by a factor of $1.5$ from the bin boundaries of \S\ref{sec:med-spec-3d}.
The change in the LOF score across this bin boundary increased moderately to LOF(bin 3) - LOF(bin 4) $= 0.130$ for $k=70$.
Thus, the plane in step (8) of our procedure
passes though a point $(0.50, 0.72, 0.75)$ in the \emph{normalised} space of
($i-W3$, $\log_{10}($REW(\civ)$)$, kt$_{80}$(\civ)) with a normal vector of
  $\hat{n}=(0.64, 0.41, 0.64)$. We thus define T1BERQs by the following inequalities
  in the 3D \emph{normalised} parameter space:
 \begin{align}
  \label{eq1:3d-boundary}
  0.64(i-W3) + &0.41\log_{10}REW + 0.64kt_{80}(\civ)-1.10\ge 0.\\
  \label{eq2:3d-boundary}
  \theta &\le 19.6^{\degr},
 \end{align}
where $\theta$ is defined in Eq. \ref{theta}.

Figure~\ref{fig:3d-boundary} visualises the resulting set of quasars, T1BERQs, in 2D projections of the 3D space. Quasars are colour coded to show those which would be selected by both the T1CERQ and the T1BERQ criteria, by one but not the other, or by neither. Quasars selected by T1CERQ but not T1BERQ (15 quasars) are those with low kt$_{80}$: they are red and possess strong but not boxy \civ~lines. Quasars selected by T1BERQ but not T1CERQ (76 quasars) are those which our local outlier factor selection algorithm judged to be closer to the ERQ subset than the main quasar locus. They are generally somewhat less red than the other ERQs and have weaker \civ~lines, but exhibit the same extreme line properties. Overall, the T1BERQ selection produces 133 quasars.

\begin{figure*}
  \includegraphics[width=\linewidth]{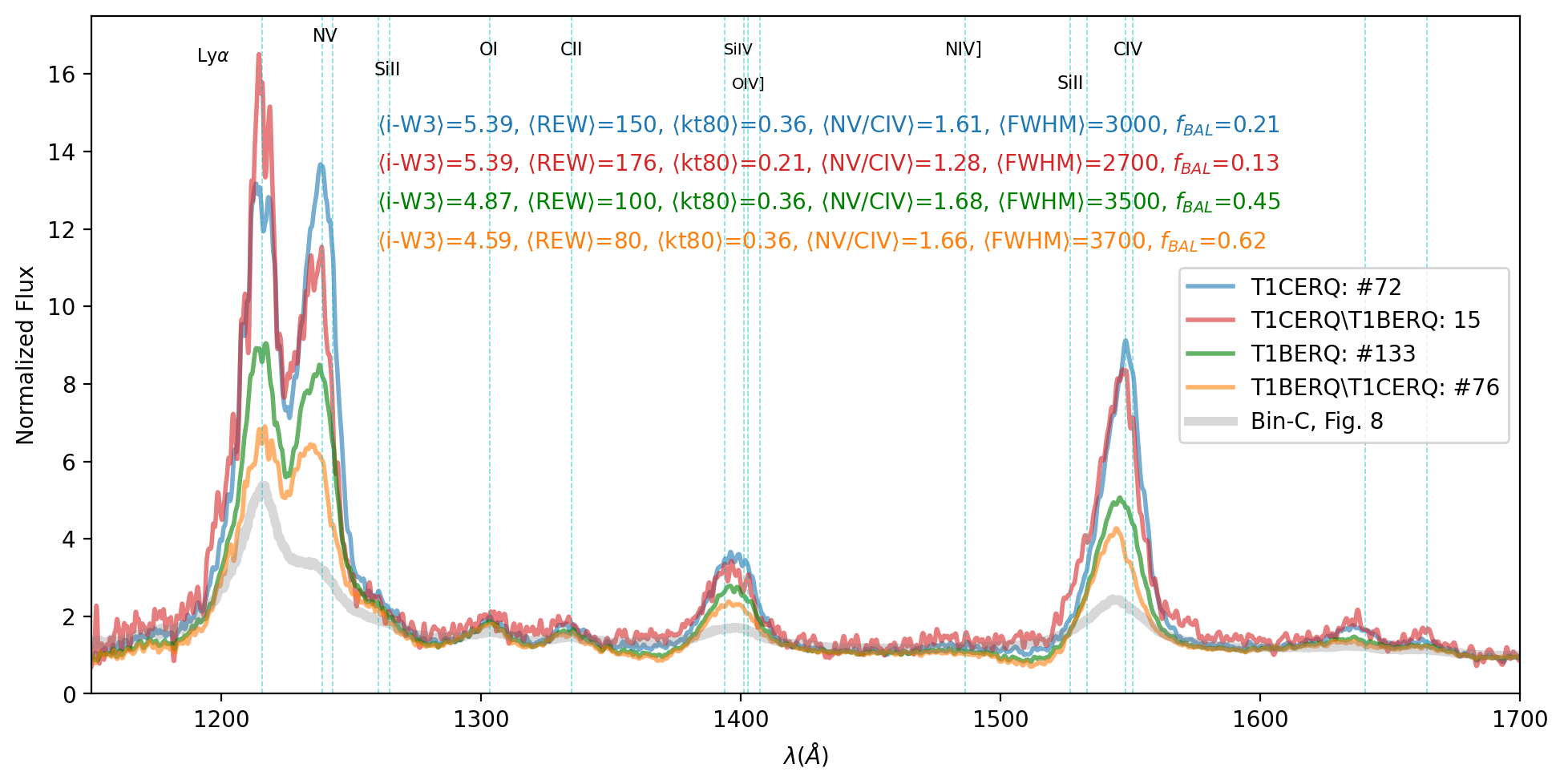}
  \caption{The median spectrum of T1CERQs is shown by the blue curve. The median spectrum
   of T1CERQs which are not among T1BERQs is plotted with the red curve.
    The median spectrum of those T1BERQs which are not among T1CERQs is shown in orange.
    The thick grey curve shows the median spectrum of all quasars
     in the dense region within  bin C in Figure \ref{fig:3d-bin}.}
  \label{fig:med-spec-BERQ-CERQ}
\end{figure*}

\begin{figure}
  \includegraphics[width=\linewidth]{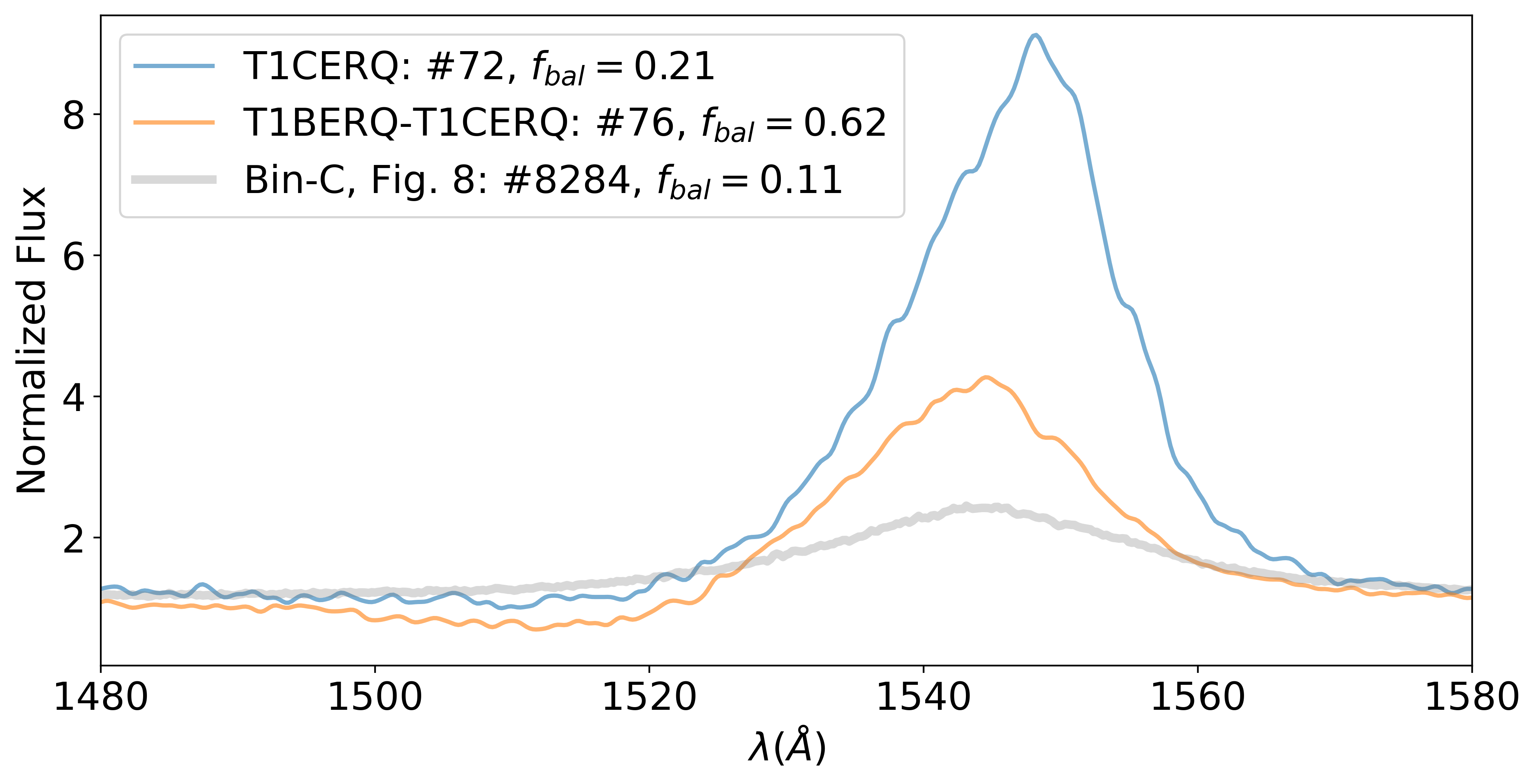}
  \caption{This figure shows a zoom in view of the median spectra around \ion{C}{IV} BAL region.
  Our newly classified objects (i.e. T1BERQs which are not among T1CERQs) are compared to
  T1CERQs and to the quasars in the central bin of Fig. \ref{fig:3d-bin} and Fig. \ref{fig:3d-medspec}.
  Our 76 newly classified quasars have a higher visually verified BAL fraction.  }
  \label{fig:inset}
\end{figure}

Figure \ref{fig:med-spec-BERQ-CERQ} compares the median spectra of T1BERQs to
T1CERQs. As expected given the selection criterion, the T1BERQ sets have 
higher average kt$_{80}$, but lower average $i-W3$ and lower REW(\civ).
However, they also exhibit the other unusual line properties associated
with T1CERQs, to a stronger extent. In particular, the 76 quasars in 
T1BERQs but not T1CERQs have a high BAL fraction of $f_{BAL}=0.62$, roughly 
three times larger than T1CERQs (see Fig. \ref{fig:inset} for a clearer comparison). 
The low kt$_{80}$ quasars which were removed
were also those with the lowest BAL fraction. The FWHM of the \civ~line is 
larger in the newly selected T1BERQs, strengthening the general trend shown 
in Table~\ref{tab-3d-properties}. Finally the \ion{N}{V} line is strong, as
shown by the high \fnc, and visually in the median
spectra, where \ion{N}{V} strength is comparable to the Lyman-$\alpha$ emission
line. 

\subsection{T1BERQs in WISE AGN catalogue}
\label{sec:milliquas}
To determine whether our sample of T1BERQs are extremely red only within the
SDSS colour selection criteria or are extreme also as a part of other quasar samples, we performed
a parallel analysis using the MILLIQUAS catalogue \citep{MILLIQUAS}.
We cross-matched the quasars in MILLIQUAS
with the WISE AGN catalogue \citep{WISEA}, as the 
infrared flux measurements of WISE are well-suited 
to studying red quasars like ERQs (H17). MILLIQUAS 
is a compendium of extant spectra with a high 
likelihood of being quasars. As SDSS is the
largest spectral survey in existence, most, but
not all, spectra in MILLIQUAS come from SDSS. 
If the colour selection function of SDSS were
truncating the ERQ distribution, we would expect 
that the set of objects in MILLIQUAS but not in 
SDSS would extend substantially further towards
the locus of ERQs in WISE colour space.

\begin{figure}
  \includegraphics[width=\linewidth]{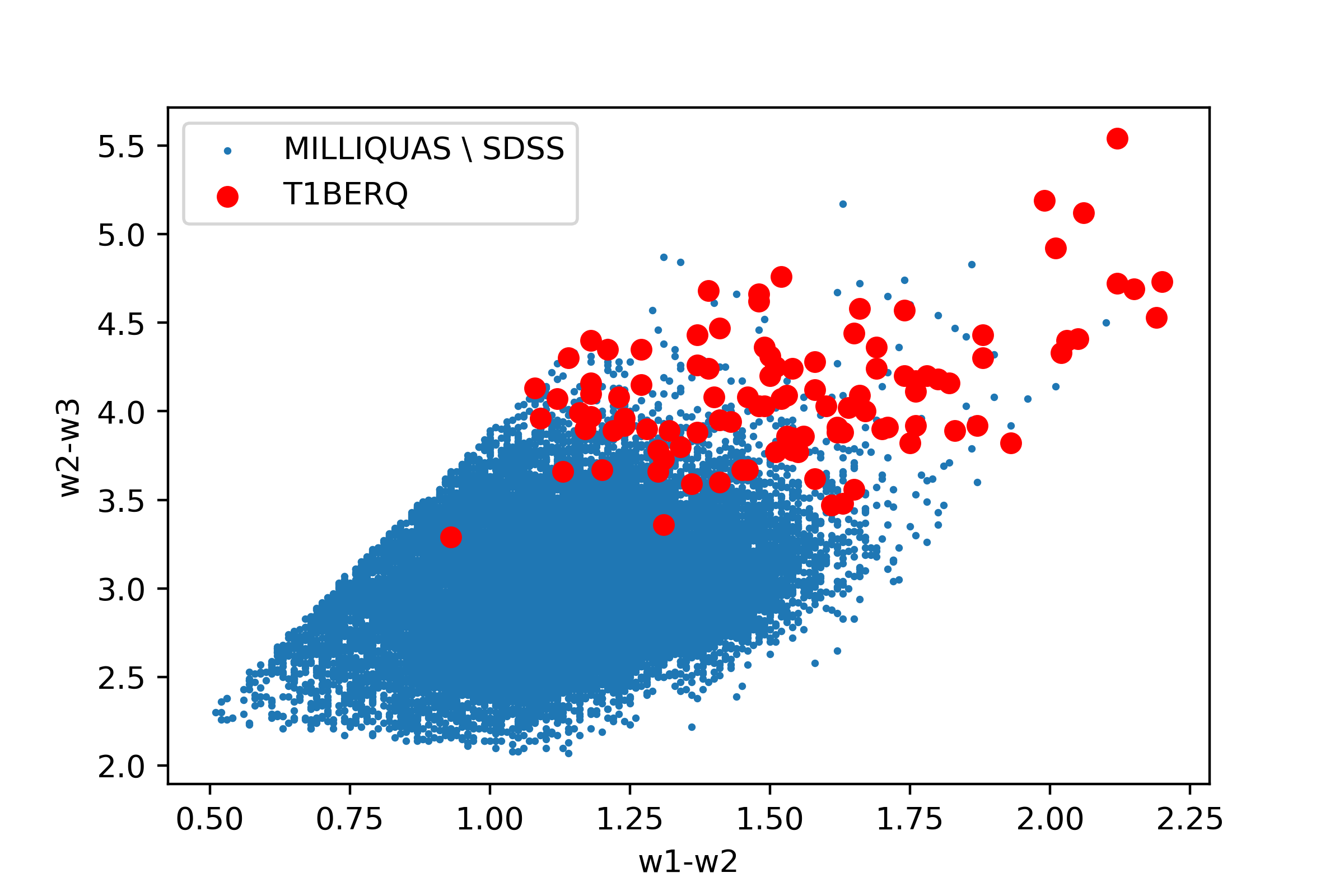}
  \includegraphics[width=\linewidth]{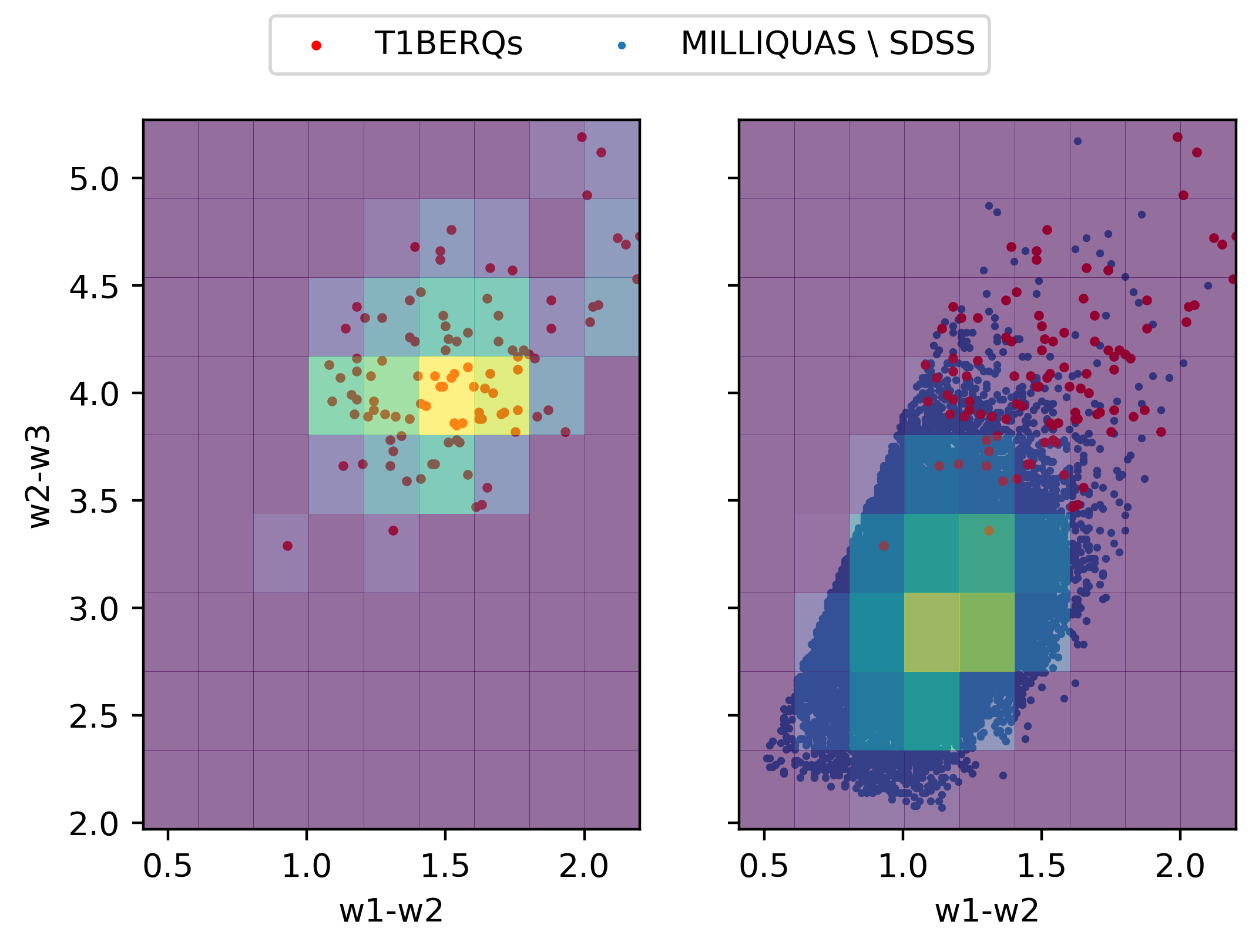}
  \caption{(Upper): blue dots are all quasars in the  MILLIQUAS  catalogue,
           but not in the SDSS catalog, with a spectroscopic redshift.
          Red dots are T1BERQs. 
          (lower left) 2D histogram of T1BERQs in this colour space.
        (Lower right): 2D histogram of all quasars in the upper panel, with T1BERQs shown as red dots.
          }
  \label{fig:2dhist}
\end{figure}

Figure~\ref{fig:2dhist} shows our selected
sample of T1BERQs in the
($w1-w2, w2-w3$) colour-colour space\footnote{Unfortunately, the $i-$band magnitudes for the quasars in our comparison
sample are not available; otherwise it would be very 
illustrative to  compare T1BERQs with our comparison 
sample in a 3D colour-colour-colour plot of $(i-w3, w1-w2, w2-w3)$.} of the WISE catalogue. 
We also show our comparison sample (i.e. crossmatched WISE AGN 
and MILLIQUAS quasars that have spectroscopic redshift but are
not listed in SDSS). MILLIQUAS does 
extend the quasar locus moderately
towards low $w2-w3$, but this is in the
opposite direction to the T1BERQs. 
There is no evidence that colour selection
effects are skewing our sample.
We also show the histogram of T1BERQs in
colour-colour space, as well as the 
histogram of the MILLIQUAS catalogue. 
These two histograms clearly have separate
centers, indicating that even though T1BERQs are
originally selected in the SDSS, they are extreme 
even in the MILLIQUAS  catalogue \emph{after excluding SDSS
quasars.}

\section{Conclusions}\label{sec:conclusion}
We have studied the phenomenon of extremely red quasars (ERQs),
found by \cite{hamann17} (H17) to have red colour, large REW(\civ) and
unusual emission line properties, including boxy \civ \ line profiles
a high incidence of blue-shifted BALs, and
high [OIII] 5007\AA \ outflow speeds up to 6702 km s$^{-1}$ \cite{serena19}.
These properties are consistent with ERQs being consistent with an early dusty
stage of quasar-galaxy evolution, where strong quasar-driven
outflows provide important feedback to the host galaxies. In this paper,
we have used data driven techniques to understand whether the ERQs, when
mapped into spectral parameter space, represent a separate population
and, if so, where the boundaries of this population lie.

We applied a kernel density estimation in the space of $i-W3$ colour
and \civ~rest equivalent width identified by H17 to show that the
ERQs produce overdensities. We computed the local outlier
factor, previously calibrated on mock data, to assess whether
these overdensities could be explained as statistical
fluctuations at large distance from the median sampled quasar. 
The signature of two separate populations, as we showed using
mock data, is a dip in the local outlier factor at the boundary
between the populations. In two dimensions there was a dip
near the boundary of the ERQs, but it was not strong enough
to rule out a statistical fluctuation. We therefore
considered higher dimensionality
space, adding a third parameter, kt$_{80}$, defined by H17 as
a measurement of line kurtosis or boxiness, and correlated
with the presence of an ERQ. In the three dimensional
space of $i-W3$ colour, \civ~rest equivalent width
and kt$_{80}$, we found a strong dip in the
local outlier factor around the boundaries of
ERQs, with a cluster size of $100$-$150$. The dip in the local 
outlier factor provides evidence that ERQs are connected to a
distinct phase of quasar formation, rather than being part of
a smooth transition from normal blue quasars to the tail of colour-REW distribution 
towards redder colours and larger REW(\civ)s.

We refined the selection criteria for T1CERQs, resulting in a new sample of `boxy' ERQs (T1BERQs). The idea behind these selection criteria is to use line emission properties to better align the boundary of the T1CERQ sample with the onset of the special phase of quasar formation that leads to these exotic quasar properties. To do this, we made use of the `boxy' shape of the \ion{C}{IV} line, defining a boundary in 3D which maximised the depth of the dip in the LOF score. Our final sample defined T1BERQs by the inequalities \ref{eq1:3d-boundary} and \ref{eq2:3d-boundary}, which refer to the  common region between a plane and a cone obtained by finding the largest minimum of an LOF score in a bin.

There are 15 quasars in the sample of T1CERQs which are not in T1BERQs.
Despite having very red colour and extremely strong \civ \ lines, these quasars
have much lower kt$_{80}$(\civ) compared to the average T1CERQ and are thus excluded on the basis of their non-boxy line profile. On the other hand, there are 76 quasars which are within the T1BERQ sample, but are not T1CERQs. Selected on the basis of their kt$_{80}$(\civ) as well as their red $i-W3$ and high REW(\civ), these quasars have more extreme spectral properties, exclusive of the selection criteria, than the T1LM sample.
Our T1BERQ selection criteria produced \ion{N}{V} lines which were strong compared to the \civ~and Lyman-$\alpha$, and \civ~lines with a greater FWHM than expected for the quasars' colour. The T1BERQs also had a high BAL fraction of $f_{BAL}=0.62$, roughly three times larger than T1CERQs. If ERQs are associated with an early dusty stage of quasar formation, we would expect strong metal lines and a high fraction of BAL, associated with a dense accretion disc. The final result of our paper is thus improved selection criteria which produce a purer sample of these interesting objects.
This will help to identify ERQs more efficiently in up-coming large quasar
surveys such as the Dark Energy Spectroscopic Instrument (DESI) \cite{DESI} or HETDEX \cite{HETDEX},
and select the best targets for follow-up observations
investigating quasar and galaxy evolution.

\section*{Data Availability}

Our underlying quasar line catalogue is from \cite{hamann17} and is
available as the Supplemental BOSS Emission Line Catalog\footnote{\href{https://datadryad.org/stash/dataset/doi:10.6086/D1H59V}{https://datadryad.org/stash/dataset/doi:10.6086/D1H59V}}.
Our analysis scripts and 
the sample of T1BERQs as a fits table are publicly available in \texttt{GitHub}\footnote{\href{https://github.com/rezamonadi/ExtremelyRedQuasars}{https://github.com/rezamonadi/ExtremelyRedQuasars}}. 

\section*{Acknowledgements}

We are extremely grateful to Fred Hamann for his important contribution to this
paper. We are also grateful to Serena Perrotta, Marie Wingyee Lau, Ming-Feng Ho, and Jarred Gillette
for their insightful comments and suggestions. The authors appreciate the constructive comments of the anonymous   
reviewer and we would lie to thank Joseph Mazzarella  for his guidance about using NED/iPac data bases. SB was supported by NSF grant AST-1817256.
RM thanks Fred Hamann for supporting him for part of this work from NSF grant AST-1911066.

\label{lastpage}


\end{document}